\documentclass[bibyear]{aa} 

\usepackage{graphicx}
\usepackage{txfonts}
\newcommand\be{\begin{equation}}
\newcommand\en{\end{equation}}
\newcommand{\msun}{\mbox{\rm $M_{\odot}$}}

\newcommand{\arcm}{\mbox{$^{\prime}$}}

\newcommand{\jhk}{\mbox{$JHK_{\rm s}$}~}
\newcommand{\ks}{\mbox{$K_{\rm s}$}}

\newcommand{\av}{\mbox{$A_{\rm V}$~}}

\newcommand\HII{H\,{\sc ii}~}

\usepackage{color}

\begin{document}

\title{Deep near-infrared adaptive-optics observations of a young embedded cluster at the edge of the RCW\,41 \HII region}


\author{
B. Neichel \inst{1}  \and
M. R. Samal\inst{1} \and
H. Plana \inst{2} \and
A. Zavagno\inst{1} \and 
A. Bernard\inst{1,3} \and
T. Fusco\inst{1,3} }
\institute{ 
Aix Marseille Universit\'e, CNRS, LAM (Laboratoire d'Astrophysique de Marseille) UMR 7326, 13388, Marseille, France \\ \email{benoit.neichel@lam.fr} \and
Laboratorio de Astrof\'isica Te\'orica e Observacional, Universidade Estadual de Santa Cruz, Rodovia Jorge Amado km16 45662-900 Ilh\'eus BA - Brazil \and
ONERA (Office National d’Etudes et de Recherches Aérospatiales),B.P.72, F-92322 Chatillon, France }

\date{Received ..... 2014; accepted .... 2014}

 
  \abstract
   {}
   {We investigate the star formation activity in a young star forming cluster embedded at the edge of the RCW\,41 \HII region. As a complementary goal, we aim to demonstrate the gain provided by wide-field adaptive optics (WFAO) instruments to study young clusters.}
   {We used deep, \jhk images from the newly commissioned Gemini-GeMS/GSAOI instrument, complemented with {\it Spitzer} IRAC observations, in order to study the photometric properties of the young stellar cluster. GeMS is a WFAO instrument that delivers almost diffraction-limited images over a field of $\sim$2 \arcmin across. The exquisite angular resolution allows us to reach a limiting magnitude of $J$ $\sim$ 22 for 98\% completeness. The combination of the IRAC photometry with our \jhk catalog is used to build color-color diagrams, and select young stellar object (YSO) candidates. The \jhk photometry is also used in conjunction with pre-main sequence evolutionary models to infer masses and ages. The $K$-band luminosity function is derived, and then used to build the initial mass function (IMF) of the cluster.
   }
   {We detect the presence of 80 YSO candidates. Those YSOs are used to infer the cluster age, which is found to be in the range 1 to 5 Myr. More precisely, we find that 1/3 of the YSOs are in a range between 3 to 5 Myr, while 2/3 of the YSO are $\leq$ 3 Myr. When looking at the spatial distribution of these two populations, we find evidence of a potential age gradient across the field that suggests sequential star formation. We construct the IMF and show that we can sample the mass distribution well into the brown dwarf regime (down to $\sim$ 0.01 \msun). The logarithmic mass function rises to peak at $\sim$0.3 \msun, before turning over and declining into the brown dwarf regime. The total cluster mass derived is estimated to be 78 $\pm$ 18 \msun, while the ratio derived of brown dwarfs to star is 18 $\pm$ 5 \%. When comparing it with other young clusters, we find that the IMF shape of the young cluster embedded within RCW\,41 is consistent with those of Trapezium, IC\,348, or Chamaeleon I, except for the IMF peak, which happens to be at higher mass. This characteristic is also seen in clusters like NGC\,6611 or even Taurus. These results suggest that the medium-to-low mass end of the IMF possibly depends on environment.}
   {}

   \keywords{Stars: formation -- circumstellar matter -- ISM: bubbles -- \HII regions -- Infrared: stars -- Instrumentation: adaptive optics -- Instrumentation: high angular resolution
                 }

\titlerunning{Young NIR cluster embedded in RCW\,41}

   \maketitle
%

\section{Introduction}
Massive stars form \HII regions that expand in the surrounding medium.  At supersonic speed, this
expansion can create a layer of cold neutral material that is accumulated between the ionized and the shock fronts. This layer may become unstable and then collapse into fragments that form a new generation of stars through different physical mechanisms (see Deharveng et al.  \cite{deh10}), including the collect and collapse processes (Elmegreen \& Lada, \cite{elm77}) or small and large scale instabilities. Observations in our Galaxy suggest a possible causal link between multiple generations of stars observed in star forming regions (e.g., Thompson et al. \cite{tho12}, Preibisch et al. \cite{pre12}, Dirienzo et al. \cite{dir12}, Deharveng et al. \cite{deh12}, Jose et al. \cite{jos13}, Samal et al. \cite{sam14}). Interpreting these observations as evidence of triggered star formation requires detailed simulations, including all the possible sources of feedback, and several studies are currently exploring this aspect of star formation (e.g., Haworth et al. \cite{haw14}, Dale et al. \cite{dal14}, Gritschneder et al. \cite{gri14}, Walch S. K. \cite{wal14}).

On the other hand, the observational properties of clusters that formed at the edge of \HII regions are poorly known (e.g., Krumoltz et al. \cite{kru14}, Tan et al. \cite{tan14}), mainly owing to limitations in angular resolution and sensitivity. For instance, the collapse of those massive fragments may host the formation of high-mass stars, as suggested by recent statistical studies, showing that about 30\% of condensations observed at the edges of \HII regions do host massive stars (Deharveng et al.  \cite{deh10}, Thompson et al. \cite{tho12}, Simpson et al. \cite{sim12}). On the other side of the mass range, young clusters contain a large number of low-mass stars, protostellar objects, and brown dwarfs. As predicted by models, those protostellar objects and brown dwarfs will be 2 to 3 mag brighter in young clusters than in field brown dwarfs of the same mass.
Young clusters are then ideal targets for studying the nature and mass distribution of young stellar objects (YSOs), including brown dwarfs, and for understanding the physical process underlying their formation (e.g., Muench et al. \cite{mue03}, Levine et al. \cite{lev06}, Harayama et al. \cite{har08}, Oliveira et al. \cite{oli08}, Ojha et al. \cite{ojh09}).

Young clusters usually have a heavily obscured and dense environment, hence deep high angular resolution in the near infrared (NIR) is needed to resolve the stellar populations and detect the fainter members. Using such observational techniques, young cluster members can be listed and studied in detail using color-color (CC) diagrams and evolutionary tracks, which will give their age and mass distribution. The age, mass, and spatial distribution of stars in the cluster give important information about a young cluster's evolutionary stage and put constraints on star formation models (e.g., Haisch et al. \cite{hai00}, Muench et al. \cite{mue03}, Levine et al. \cite{lev06}, Comer\'on et al. \cite{com07}, Harayama et al. \cite{har08}, Oliveira et al. \cite{oli08}, Ojha et al. \cite{ojh09}, DeRose et al. \cite{der09}, Gennaro et al. \cite{gen11}, Preibisch et al. \cite{pre11}, Rochau et al. \cite{roc11}, Santos et al. \cite{san12}, Bik et al. \cite{bik14}).



We selected a young cluster observed at the edge of the Galactic ionized region RCW~41 (Rodger, Campbell, Whiteoak \cite{rcw60}) to test how wide field adaptive optics (WFAO) can help in reaching these goals. In particular, this cluster has been chosen because it has been studied in depth before, however at seeing limited angular resolution, in the NIR (Ortiz et al. \cite{ort07}, Roman-Lopes et al. \cite{rom09}, Santos et al. \cite{san12}), and it hosts massive young stars and multiple masers emission in its vicinity as signposts of high-mass star formation (e.g., Walsh et al. \cite{wal98}, Pestalozzi et al. \cite{pes05}, Caswell et al. \cite{cas10}, Breen et al. \cite{bree11}, Voronkov et al. \cite{vor14}). Combining high-angular resolution, and deep NIR images, we are able to study the complete populations of this young star cluster, well below the hydrogen-burning limit (0.08 \msun).

\begin{figure}[ht!]
   \centering
  \begin{tabular}{c}
    \includegraphics[width= 1.1\linewidth]{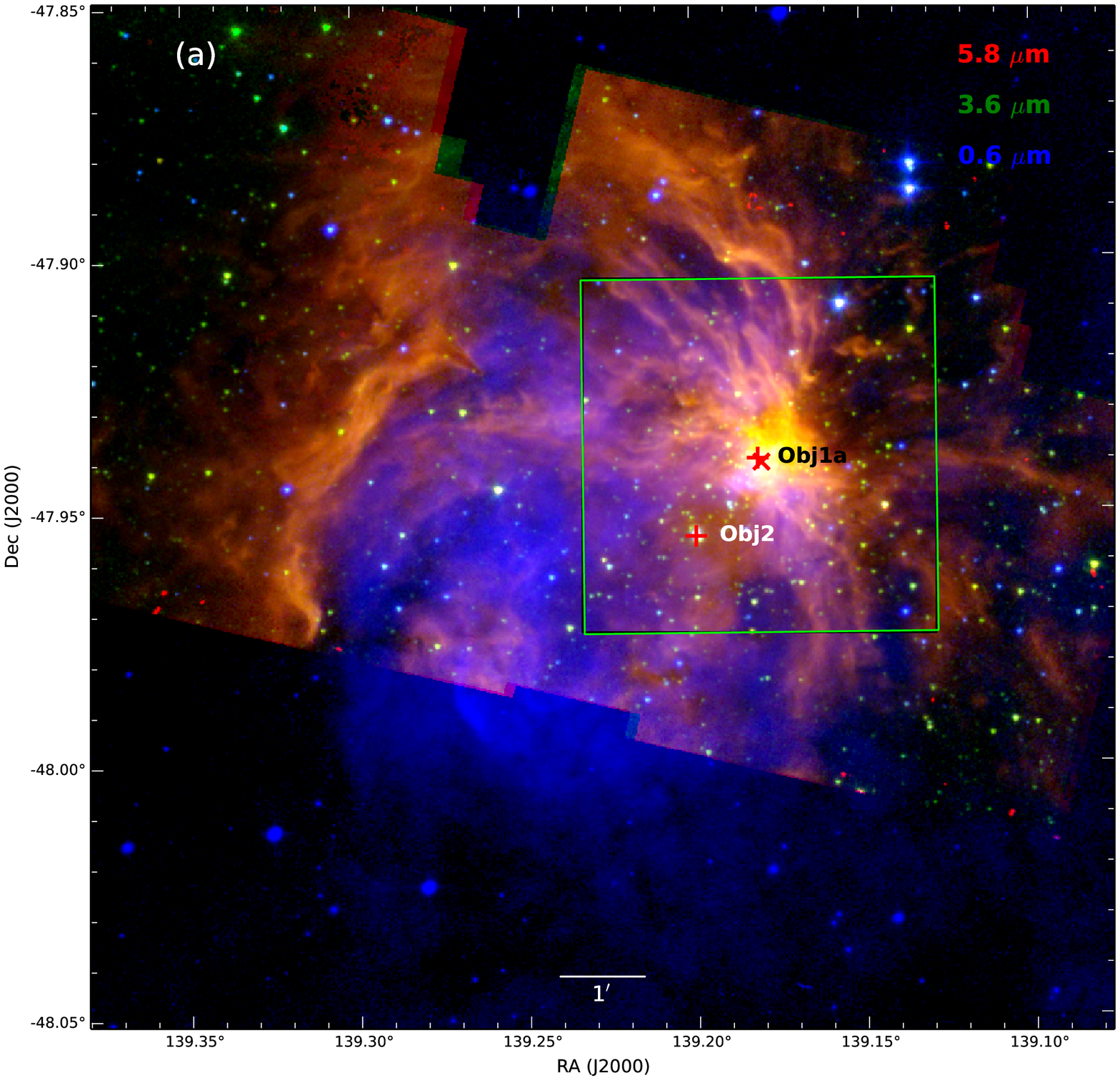} \\
  \includegraphics[width=1.1\linewidth]{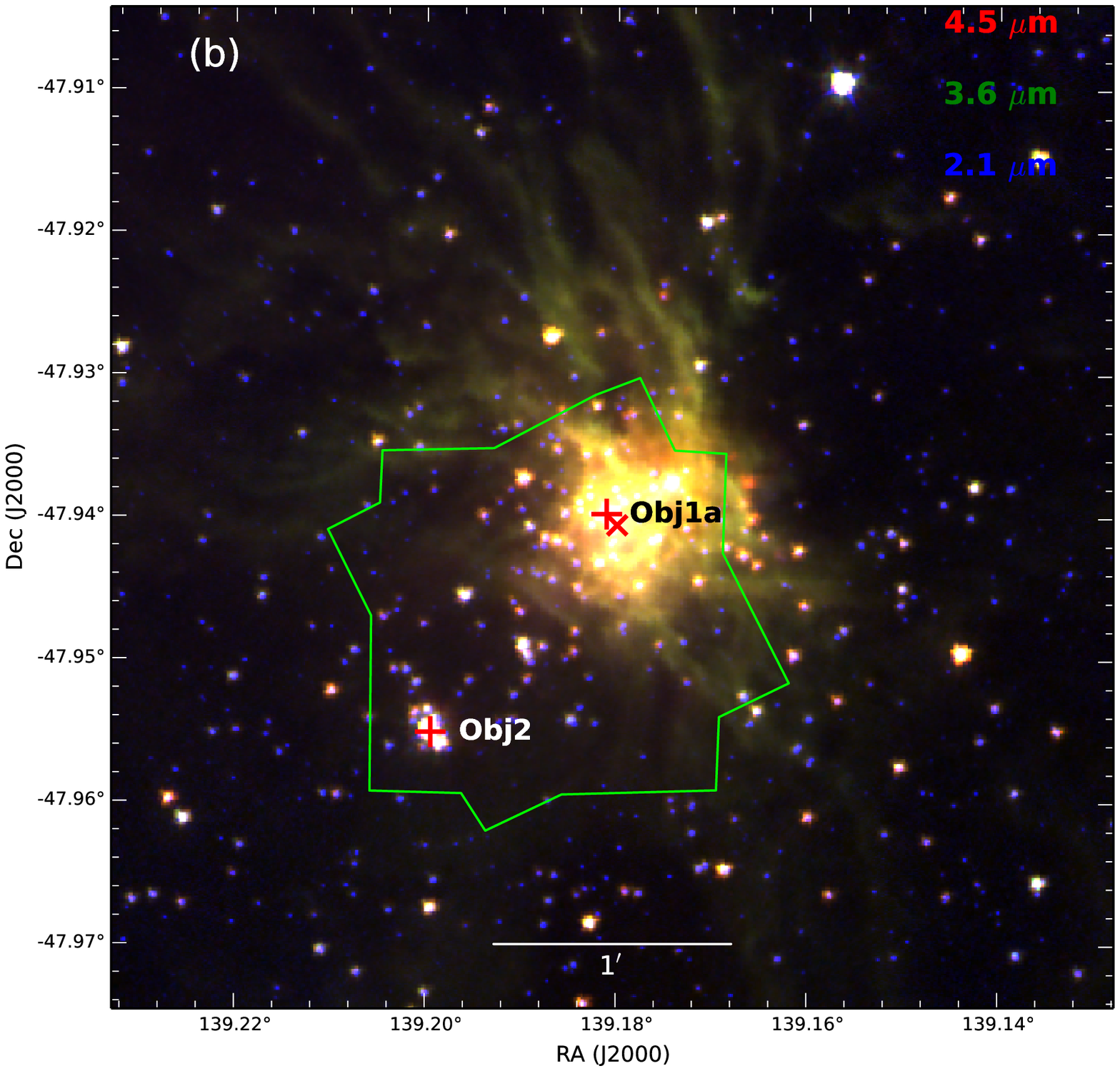}  
  \end{tabular}
   \caption{RCW~41 region. (a) H$\alpha$ (blue), 3.6 $\mu$m (green), and 5.8 $\mu$m (red) emissions. (b) 2.12 $\mu$m (blue), 3.6 $\mu$m (green), and 4.5 $\mu$m (red). The green contour in fig. (a) is the field shown in fig. (b). The green contour in fig. (b) is the field covered by GeMS/GSAOI in the present study. The red cross marks the position of the NIR source IRAS09149 - 4743.  The plus symbols show the positions of Obj1a and Obj2. North is up and east is left.}
              \label{FigH2}
\end{figure}

RCW~41 (RA = 09$^h$16$^m$44$^s$ ; DEC = -47\degr 56\arcmin 51\arcsec) is a Galactic \HII region located in the Vela molecular ridge (VMR). Figure~\ref{FigH2} (top) shows the UKST Super-COSMOS H$\alpha$ emission (Parker et al. \cite{park05}), the 3.6 $\mu$m {\it{Spitzer}} emission, and the 5.8 $\mu$m {\it{Spitzer}} emission of RCW~41. Figure~\ref{FigH2} (bottom) shows the \ks-band emission from {\it{SofI}}  (Son of Isaac, NIR camera mounted on the 3.58 meter New Technology Telescope (NTT) from La Silla/Chile - Moorwood et al. \cite{moo98}), the 3.6 $\mu$m {\it{Spitzer}} emission, and the 4.5 $\mu$m {\it{Spitzer}} emission. The GeMS/GSAOI field is shown with the contours in Figure~\ref{FigH2} (bottom). The large scale morphology of the region shows strong H$\alpha$ emission surrounded by a shell of infrared emission, which is particularly visible at 5.8 $\mu$m. The mid-infrared emission defines a ring or bubble of warm dust and polycyclic aromatic hydrocarbons (PAHs) that encompasses the ionized area. 

\begin{figure*}
   \centering
 \includegraphics[width=\hsize]{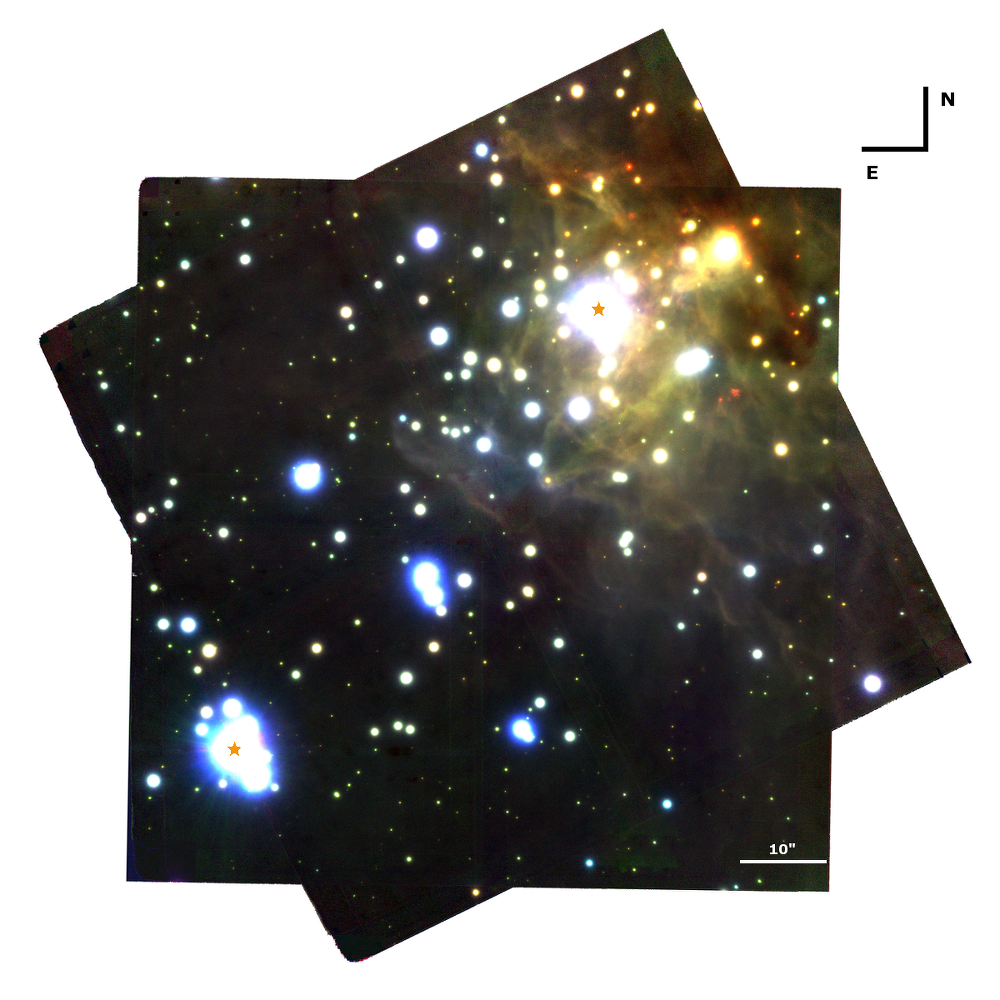}
   \caption{RCW\,41 false three-color GeMS/GSAOI image combining $J$ (blue), $H$ (green), and \ks (red). North is up and east is left. The scale is indicated in the image. The location of Obj1a (upper right) and Obj2 (bottom left), as defined in Santos et al. (\cite{san12}), are symbolized by orange stars on the image.}
              \label{fig:threecolor}%
\end{figure*}

This structure suggests that UV radiation from massive stars and high-energy photoionized electrons from inside the \HII region heats the dust grains and excites the PAHs from the surrounding environment, generating the envelope type of emission. When looking at Fig. \ref{FigH2} (top), one striking element is the presence of several conspicuous filaments directed almost radially from the cluster. This morphology has also been highlighted by Santos et al. (\cite{san12}), where they even show that the main polarization direction seems to be consistent with the direction of the associated filament.
At the edge of the H$\alpha$ emission, the {\it{Spitzer}} and {\it{SofI}} images show a dense NIR cluster, which is associated with the source IRAS$09149-4743$ (marked with a cross in both figures). This region has been studied in detail by Ortiz et al. (\cite{ort07}), Roman-Lopes et al. (\cite{rom09}), and Santos et al. (\cite{san12}). The cluster's region is composed of two distinct substructures, as highlighted in Figs. \ref{FigH2} and \ref{fig:threecolor}. The main region (called main cluster in the following), located in the northwest part of the GeMS/GSAOI field, is composed of a number of objects concentrated around a group of stars (named Obj1a/b in Santos et al. (\cite{san12})) and associated with an extended emission region. The source Obj1a (Fig. \ref{FigH2}) has been classified as a B0 V star, and it is at a distance of 1.2 $\pm$ 0.12 kpc (Roman-Lopes et al. \cite{rom09}). Pirogov et al. (\cite{pir07}) report the detection of CS, N$_2$H$^+$, and 1.2 mm dust continuum emissions toward IRAS$09149-4743$, with the dust emissions forming an almost spherical core superposed to the northwestern part of the stellar cluster direction. A small subcluster region is located 1.1 arcmin toward the southeast of the main cluster's center, and hosts Obj2 (Fig. \ref{FigH2}), identified as an O9 v star at a distance of 1.27 $\pm$ 0.13 kpc (Roman-Lopes et al. \cite{rom09}), which is presumably the most massive object in the field. Following the conclusions drawn in Santos et al. (\cite{san12}), we adopt a cluster distance of 1.3 $\pm$ 0.2 kpc in the rest of this paper.

We present images of the RCW\,41 cluster corrected for atmospheric distortions using the Gemini South Muti-Conjugate Adaptive Optics System (GeMS - Rigaut et al. \cite{rig14}, Neichel et al. \cite{nei14a}) and obtained with the Gemini South Adaptive Optics Imager (GSAOI - McGregor et al. \cite{mcg04}, Carrasco et al. \cite{car12}). 
GeMS is a facility instrument for the Gemini-South telescope. It delivers a uniform, almost diffraction-limited image quality at near-infrared wavelengths over an extended FoV of 2 arcmin across. GeMS uses five artificial laser guide stars, up to three natural guide stars, and multiple deformable mirrors (DMs) that are optically conjugated with the main turbulence layers, resulting in a AO corrected field that is 10 to 20 times larger than previous AO systems (also called single-conjugate AO - SCAO). The GSAOI focal plane is formed by a 2 $\times$ 2 mosaic of Hawaii-2RG 2048 $\times$ 2048 pixel arrays with 3.0" wide gaps. Images are recorded in a 85" $\times$ 85" field of view with a plate scale of $\sim$20 mas. 
GeMS/GSAOI observations provide a far deeper and sharper view of the cluster than previously available. These new images allow us to reassess the stellar contents of the cluster, identify YSOs  among its members, provide a detailed analysis of the mass function, and provide further insight into the low-mass content of RCW\,41. In Sect. \ref{sec:datared} we present the data. Section \ref{sec:datared2} describes the specific data reduction tools that were developed for this analysis. Section \ref{sec:compcont} addresses the completeness and contamination of the data. Section \ref{sec:cd} discusses the CC and color-magnitude (CM) diagrams. In Sect. \ref{sec:imf} we derive the $K$-band luminosity function (KLF), and associated initial mass function (IMF) of the cluster. Section \ref{sec:discu} discusses the results and Section \ref{sec:conclu} gives the conclusions.

\section{Presentation of the data}
\label{sec:datared}
\subsection{GeMS/GSAOI data}

The data were obtained between January 27and February 2, 2013 as part of program GS-2013B-SV-413 (P.I. H. Plana). This program is one of the 13 programs that were selected for the GeMS/GSAOI science verification campaign. The details of the observations are summarized in Table~\ref{tab1}. Figure \ref{fig:threecolor} shows the final three-color image built from the GeMS/GSAOI data. Images were recorded through $J$, $H,$ and \ks filters, with short- and long-exposure times to allow a wider magnitude range to be sampled than is possible with only a single exposure time. Each observation consists of four science fields (each one dithered by 4" to remove gaps between the detectors) and four adjacent sky frames taken 2.5\arcmin  away from the science field. Those sky fields will be used as control fields. The averaged resolution over the field, measured as the FWHM of the stars on single-exposure frames, is reported in Table~\ref{tab1} as is the natural seeing. The averaged FWHM in \ks is 95 mas, while the averaged Strehl ratio (SR) is 14\%. This is a typical performance of GeMS/GSAOI under suboptimal natural seeing conditions (reported as IQ85 condition in Neichel et al. (\cite{nei14a})). But since RCW\,41 is not a very dense cluster, those IQ85 conditions were acceptable for this program. Overall, we get an improvement of the resolution by a factor of 4 to 6 over natural seeing and over previous NIR data sets (Santos et al. \cite{san12}).

\subsection{Wide field AO performance}
From the system performance point of view, what is interesting in Table~\ref{tab1} is to compare the performance during the nights of 27 and 30. For instance, for the \ks filter, the SR has been improved by a factor almost 3 between those two nights, while the natural seeing conditions were almost identical (1.1 \arcsec for the 27, 0.95 \arcsec for the 30). As described in Vidal et al. (\cite{vid13}), for both nights similar MCAO loops settings were used, so the difference cannot be explained by instrumental adjustments. In fact, the difference in performance can be explained by the difference in the vertical distribution of the turbulence over those two nights. Even though the integrated turbulence strength is approximately the same, for an MCAO system, the performance also critically depends on how the turbulence is distributed above the telescope and on how well the main turbulent layers match the optical conjugation of the deformable mirrors (DMs). The turbulence profile, also called $C_n^2$(h), can be evaluated from the MCAO loops real-time data, following the method described in Cortes et al. (\cite{cor12}). 

\begin{figure}
   \centering
 \includegraphics[width=\hsize]{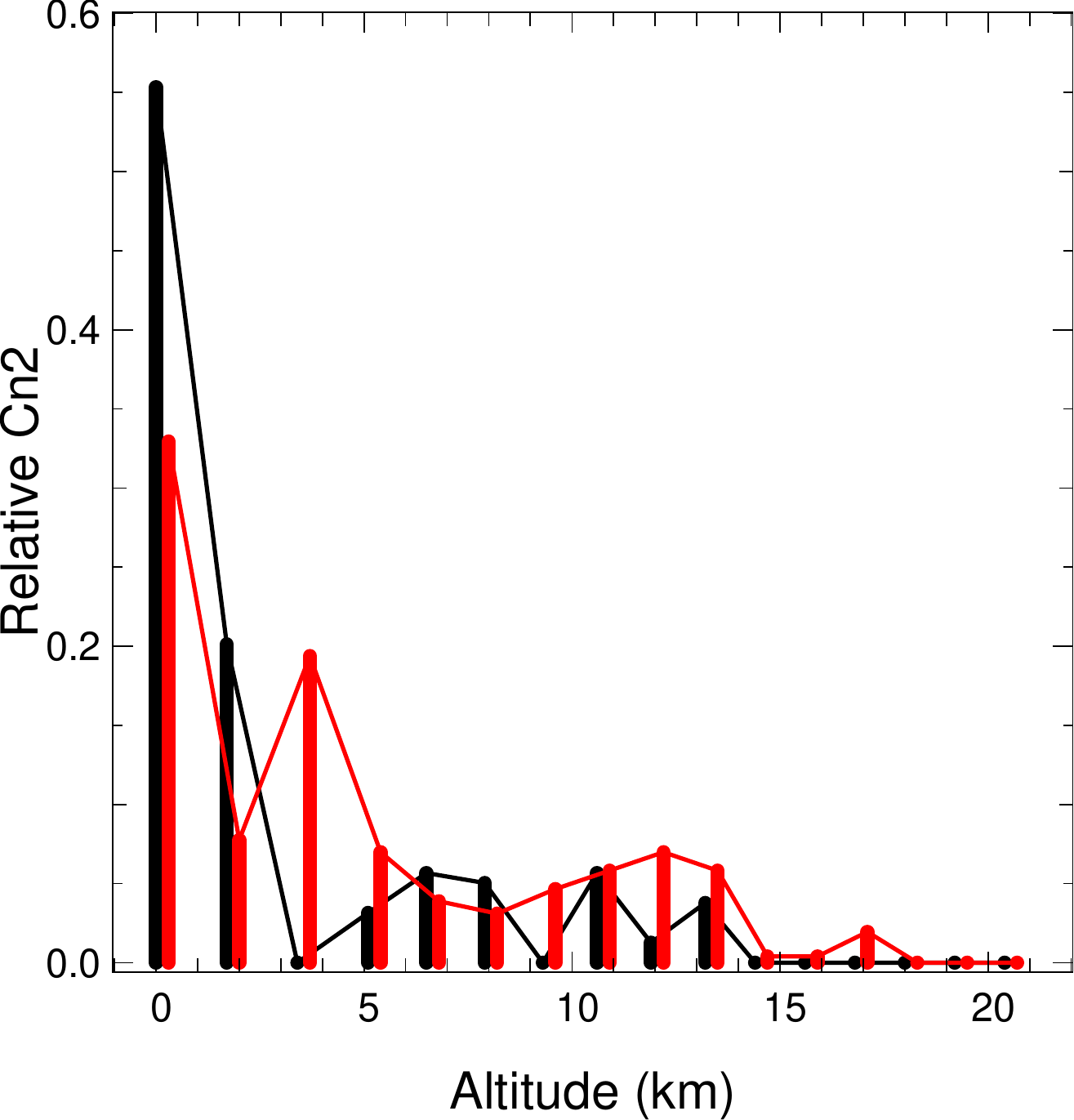}
   \caption{Vertical distribution of the turbulence (Cn2 profile), for the nights of January 27 (red) and 30 (black). The red profile has been shifted by + 200 meters for clarity.}
              \label{fig:cn2}%
\end{figure}

The normalized $C_n^2$(h) profiles for the nights of January 27 and 30 are shown in Fig. \ref{fig:cn2}. Those profiles have been normalized to one, and absolute profiles may be retrieved by scaling them by the integrated natural seeing, which in that case is almost similar for the two cases. As one can see, the vertical distribution of the turbulence differs significantly between the two nights, since the profile of the 30th is more favorable to the MCAO system, because most of the turbulence is indeed located close to the DM conjugation altitude (i.e., 0 and 9km). On the other hand, the vertical distribution of the 27 shows a strong layer located at $\sim$11km above the telescope, very likely associated with the jet stream (e.g., Guesalaga et al. \cite{gue14}), and some turbulence around 5~km, which cannot be corrected by the MCAO system. A third DM conjugated to 4.5km will be installed again in GeMS during 2015 (Neichel et al. \cite{nei14a}), which should improve the performance and the stability of the MCAO correction against different $C_n^2$(h) profiles.


\begin{table*}
\caption{Observation details. FWHMs, and SRs averaged over $\sim$ 200 stars uniformly distributed over the field and for all the individual frames. The corresponding standard deviations are also given.}         
\label{tab1} 
\begin{tabular}{lrrrrrrrrr}
\hline
\hline
Date &Filter &Individual & Number of & $<$FWHM$>$ & $\sigma_{\rm{FWHM}}$ & $<$SR$>$ & $\sigma_{\rm{SR}}$ & Natural seeing (") & PA \\
     &       & exposure time & frames  &  (mas)  & (mas) & (\%) & (\%) & (\arcsec @ 0.55$\mu$m) & (degree) \\
\hline
Jan 27$^{th}$ 2013 & $J$  & 90s & 12  & 110  & 20 & 4 & 2 & 0.65 & 27\\
                   & $H$  & 80s & 12 &  115  &  27 & 6 & 3 & 0.70 & 27\\
                   & \ks & 80s & 12  & 115  & 18 & 10 & 2 & 1.10 & 27 \\
Jan 30$^{th}$ 2013 & $J$  & 90s & 4  & 150 & 20 & 3 & 2 & 1.20 & 0  \\
                   & $J$ & 45s & 1 & 155  & 25  & 2 & 1.5 & 1.20 & 0 \\
                   & \ks & 80s & 14  & 71 & 3 & 27 & 3 & 0.95 & 0 \\
                   & \ks & 40s & 3  & 73 &  3 & 27 & 2 &  0.90 & 0\\
                  & \ks & 10s & 5  &  70 &  2 & 31 &2 & 0.75 & 0\\
Jan 31$^{th}$ 2013 & $J$  & 90s & 8 &  155 &  25 & 2.5 & 1.5 & 0.80 & 0\\
                   & $H$  & 80s & 7  &  125 & 20 & 6 & 2 & 0.75 & 0\\
Feb 02$^{th}$ 2013 & $H$  & 80s & 8 & 120 & 20 & 6 & 2 & 0.70& 0\\
                   & $H$  & 40s & 2  &  150 & 15 & 3.5 & 1.5 & 0.70 & 0\\
                   & $H$  & 10s & 10  &  135 & 10 & 5 & 2 & 0.65 & 0 \\
\hline
\end{tabular}
\end{table*}

\subsection{{\it Spitzer}-IRAC observations}
We have complemented our \jhk data with data from the Infrared Array Camera (IRAC) from the {\it Spitzer} telescope. IRAC has four wavelength bands centered at 3.6, 4.5, 5.8, and 8.0 $\mu$m, each of which having a field of view of $\sim$5\arcm.2 $\times$ 5\arcm.2. The pixel size in all the four bands is $\sim$ $1\farcs22$. The IRAC observations of the region were taken on 2007 February 17 (Program ID 30734, PI: Donald Figer). The corrected basic calibrated data (CBCD) images were downloaded from the {\it Spitzer} Space Telescope Archive using the Leopard package.
Mosaics were built with the native instrument resolution of $1\farcs22$ per pixel with the standard CBCDs using the MOPEX (Mosaicker and Point Source Extractor) software (version 18.0.1) program provided by the {\it Spitzer} Science Center. 

The area covered by our \jhk images is very bright at 5.8 and 8.0 $\mu$m owing to the diffuse emission, thus we only used the 3.6 and 4.5 $\mu$m data in the present work.
Magnitudes were extracted using the point response function (PRF) fitting method in multiframe mode and the APEX tool developed by the {\it Spitzer} Science Center. 
The standard PRF map tables\footnote{http://ssc.Spitzer.caltech.edu/irac/calibrationfiles/psfprf/prfmap.tbl} were used to fit variable PRFs across the image. Point sources with peak values of more than 5$\sigma$ above the background were considered as candidate detections. Many sources detected in the nebulosity appeared to be spurious. Those spurious sources were visually identified and deleted from the automated detection list. Following the IRAC Data Handbook, we adopted zero points of 280.9 and 179.7 Jy in the 3.6 and 4.5 $\mu$m bands, respectively.
The {\it Spitzer} detections are used in Sect. \ref{ssec:yso} to identify the NIR-excess sources of the region.

\section{GeMS/GSAOI data reduction}
\label{sec:datared2}

\subsection{Data reduction}
Data was reduced  using two independent methods. The first method used the \textsc{Iraf} tasks provided by Gemini, and the second one was based on home-made procedures developed in \textsc{Yorick} (Munro \& Dubois \cite{mun95}). The first steps are identical for
both methods: (i) creation of a master flat image based on flat images observed each night, (ii) creation of a master sky frame based on the dedicated sky images, (iii) correct the science frame from the master flat and the master sky, as well as the detector nonlinearities and different gains between each detector. At that point, the four detector frames are reduced, but not mosaicked in one extension file. This last step requires applying an instrumental distortion correction. 

For the \textsc{Iraf} method, we used a high-order distortion map derived during GeMS/GSAOI commissioning on an HST astrometric field located in the Large Magellanic Cloud (HST Proposal 10753, PI: Rosa Diaz-Miller, Cycle 14). The distortion is corrected using the program \textsc{mscsetwcs} in the \textsc{mscred} package. Once the correct World Coordinate System (WCS) is propagated, the four detector images are combined with \textsc{mscimage} in the \textsc{mscred} package. The distortion map used by the \textsc{Yorick} procedures was derived from GeMS/GSAOI observations of NGC\,288\footnote{http://www.gemini.edu/sciops/instruments/gsaoi/data-format-and-reduction?q=/node/11873} that we correlated with HST-ACS images. Both distortion maps (from \textsc{Iraf} and from \textsc{Yorick}) lead to very similar results, with a residual star-positioning accuracy of $\sim$0.2". 

This is good, but not good enough for the GeMS/GSAOI delivered image quality, which is typically around 80mas. Moreover, this distortion calibration only accounts for the static instrumental distortion. As explained in Neichel et al. (\cite{nei14b}), a dynamical distortion component, which depends on the NGS constellation and environmental factors like the telescope pointing, is added to the final distortion budget. This dynamical distortion contribution has to be corrected frame by frame, taking one of them as the astrometric reference. Note that ot applying this second step could lead to degradinh the resolution on the combined image. With the \textsc{Iraf} data reduction package, the frame cross-registration is done using the program \textsc{geomap} implemented in the \textit{imcoadd} package.
For each image, the resulting fit have an RMS of less than 0.3 pix in both axis, which is good enough to ensure no resolution degradation.
The \textsc{Yorick} procedure follow the steps described in Neichel et al. (\cite{nei14b}). High-order polynomials (15 degrees of freedom per axis) are used to cross-register the images together. This typically leads to a precision of 0.1 pixel or better. 

\begin{figure*}
   \centering
  \begin{tabular}{ccc}
  \includegraphics[width= 0.3\linewidth]{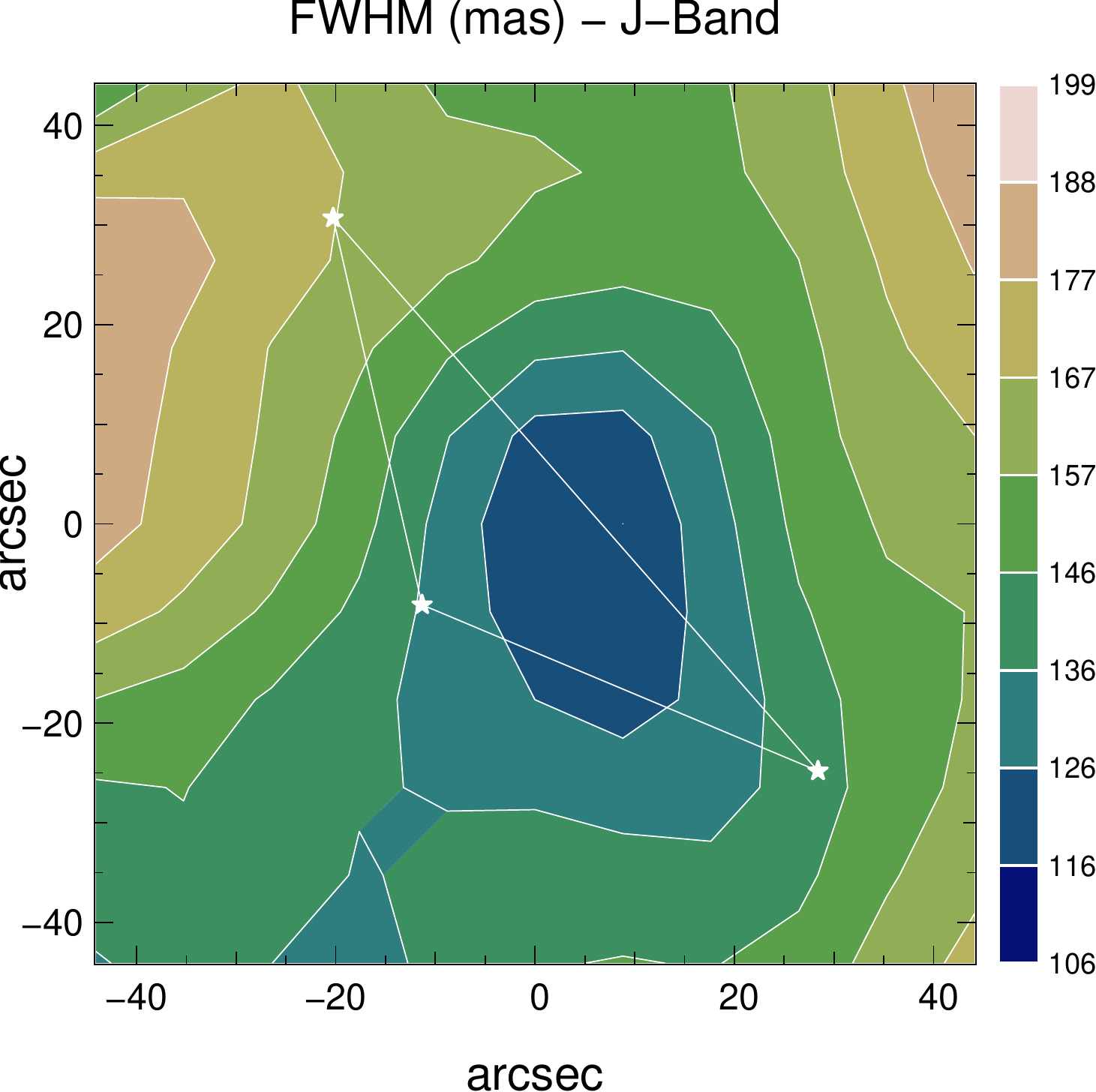} &
  \includegraphics[width=0.3\linewidth]{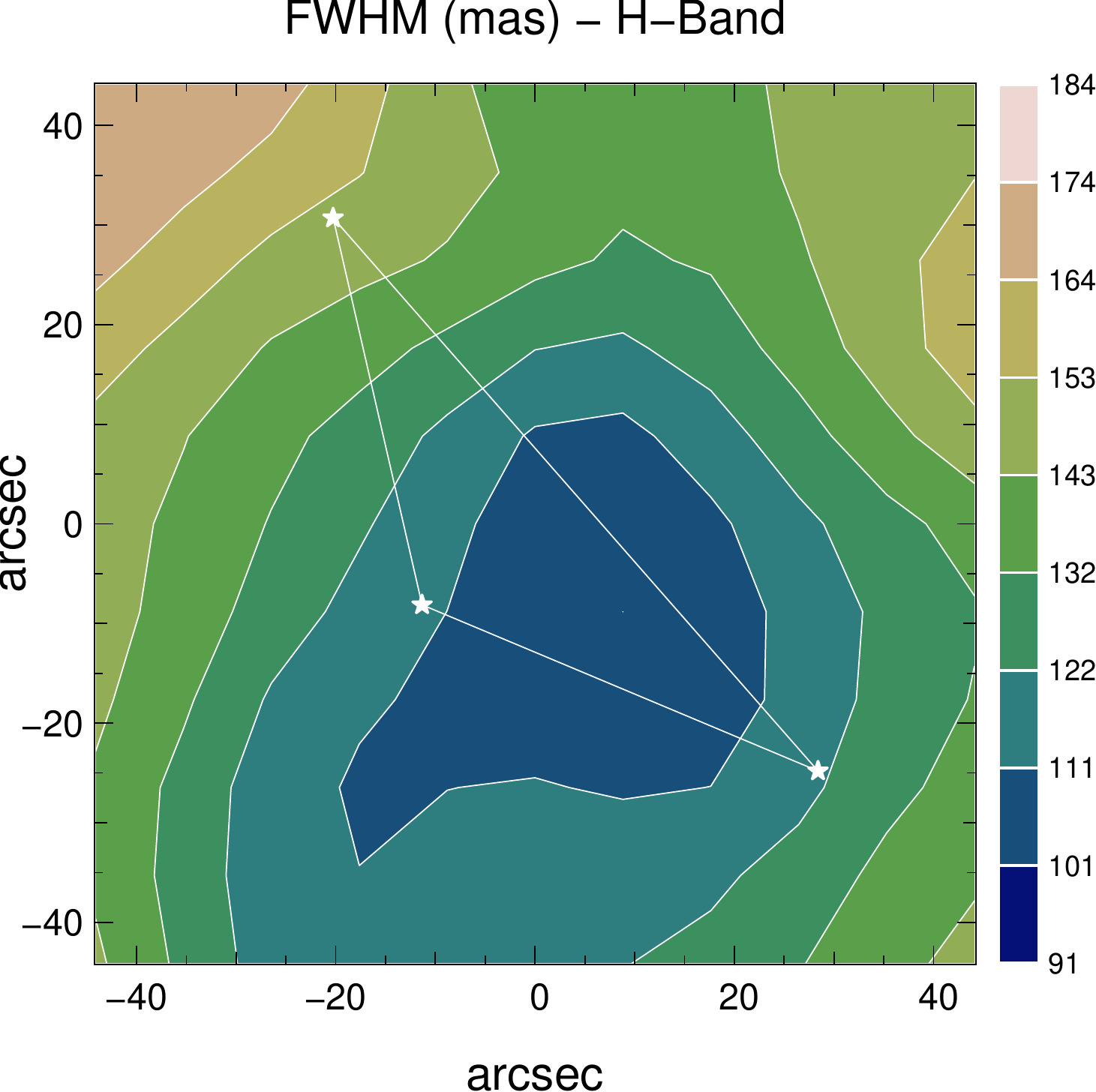}  &
    \includegraphics[width=0.3\linewidth]{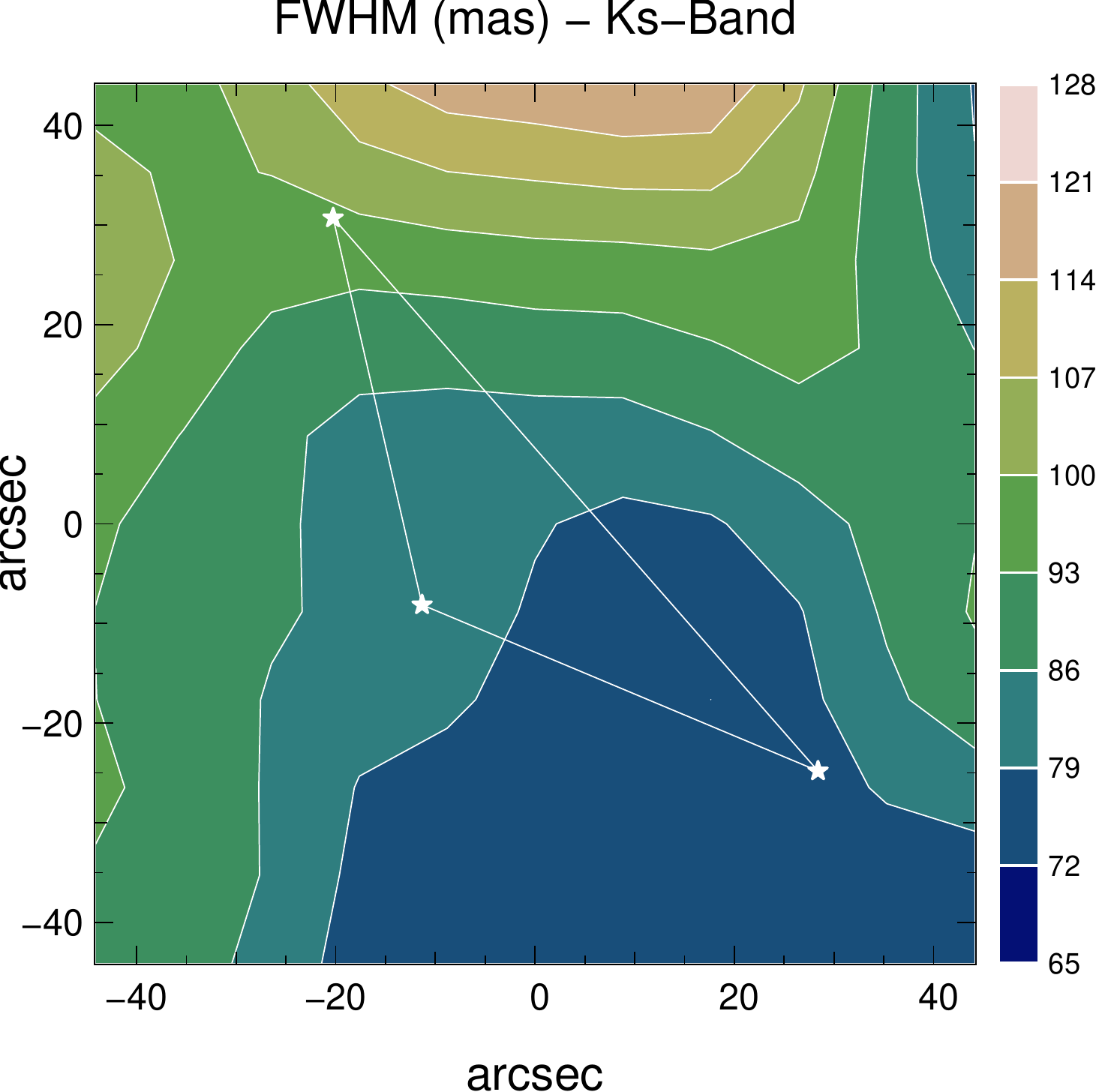}  
  \end{tabular}
   \caption{FWHM on the reduced images. The field center is at RA = 09$^h$16$^m$44.862$^s$; DEC = -47\degr 56\arcmin 51.11\arcsec. Left is $J$, middle is $H$, right is \ks. The white stars show the NGS constellation used. Data obtained on the 27th only used the 2 bottom NGS, which explains the loss in performance toward the north of the field.}
              \label{fig:fwhm}
\end{figure*}

Following the procedure described above, each individual image was reduced and eventually combined by filter to produce three long-exposure and three short-exposure reduced images. Both the total exposure time and the final resolution of the long-exposure images are summarized in Table$~$\ref{tab2}. FWHM maps for the three bands are shown in Fig. \ref{fig:fwhm} and a false three-color image built with the long-exposure individual image is shown in Fig. \ref{fig:threecolor}. At the cluster distance (1.3 kpc), the field covered by GSAOI represents $\sim$0.5 pc, with a resolution that corresponds to 130 AU. Looking at Fig. \ref{fig:fwhm}, one can clearly see a gradient in the performance across the field, the southern part getting better correction than the northern part of the field. Because of a technical problem, this is actually due to all data taken on the night of the 27th having been acquired with only two NGS (the two bottom ones in Fig. \ref{fig:fwhm}), so the north part of the field was not properly covered and corrected for low-order aberrations. As a matter of fact, stars are more elongated owing to anisoplanatism on that region of the field. The effect is somehow compensated for by the data acquired the other nights, using all three NGS, thereby providing better coverage of a field and better uniformity of the PSFs. The impact of PSFs variations on photometry accuracy is discussed in Sect. \ref{ssec:zp}.  

Finally, when doing a pixel-by-pixel comparison of the reduced and combined images obtained from both reduction methods (\textsc{Iraf} vs. \textsc{Yorick}), no bias is found, and the RMS of the difference is lower than 5\%. We therefore conclude that both data reduction processes are validated and that no major impact on the photometry analysis is to be expected from the data reduction pipeline. In the following, only the \textsc{Yorick} images are used.

\begin{table}
\caption{Reduced images characteristics. FWHMs and SRs are averaged over $\sim$ 200 stars uniformly distributed over the field. Corresponding standard deviations are also given.}            
\label{tab2} 
\begin{tabular}{cccccc}
\hline
\hline
Filter & Total &  $<$FWHM$>$ & $\sigma_{\rm{FWHM}}$ & $<$SR$>$ & $\sigma_{\rm{SR}}$ \\
         &   exposure time &   &  & & \\
\hline
$J$ & 2160 s. & 150 & 30 & 2 & 1 \\
$H$ & 2160 s. & 140 & 30 & 5 & 1.5 \\
\ks & 2080 s. & 95 & 25 & 14 & 2 \\
\hline
\end{tabular}
\end{table}

\subsection{Astrometry}
As stated above, GeMS/GSAOI images suffer from a significant field distortion, mostly owing to the instrument optical relay that consists of two off-axis parabolic mirrors. These distortions can be compensated for when co-adding the images; however, the absolute WCS need to be calibrate with an external catalog. Astrometry with GSAOI is nontrivial because there is a large discrepancy between the angular resolutions of GSAOI and common all-sky astrometric reference catalogs. Several sources taken from astrometric reference catalogs are actually found to be multiple systems when observed with GSAOI. For that reason, we chose to use the Santos et al. (\cite{san12}) catalog, because it was built with the highest, previously available angular resolution. We picked a set of 100 stars from the Santos et al. (\cite{san12}) catalog, which is uniformly distributed over the GSAOI field and for which the GSAOI images show no close companion. The WCS parameters are then computed by fitting the GSAOI X/Y star positions, with the RA/DEC sky positions taken from the catalog. Only linear transformations (shift, scale, and rotation) are applied at that point, which leads to an averaged astrometric error of $~$80mas (maximum difference being 200mas). The average precision is less than the averaged FWHM of the GSAOI images, so no confusion is expected.

\subsection{Flux measurement and zero-point calibration}
\label{ssec:zp}

Flux measurement used the following two distinct methods, one with \textsc{DAOPhot} (Stetson et al. \cite{ste87}) and the other based on \textsc{StarFinder} (Diolaiti et al. \cite{dio00}). Before describing the steps followed in both paths, one important point to emphasize is that the AO PSF is fundamentally different from classical seeing-limited PSFs. AO PSFs usually show a complex structure, which combines a close-to-diffraction core, on top of rather extended wings (see Neichel et al. (\cite{nei14a}) for a PSF model for GeMS). The complete PSFs then has a diameter that is approximately the same as the seeing disk, even if a large amount of the flux is concentrated inside an angular region of at most a few times the diffraction limit. For this reason, AO-photometry requires PSF-fitting algorithms, which provide more accurate results than regular aperture photometry. Moreover, the AO-PSF is not constant over the field or between frames. And even if MCAO instruments, such as GeMS, significantly improves the uniformity of the AO-correction over the FoV when compared to single-conjugate AO (SCAO), it remains difficult to predict the PSF variations across the FoV. As a result, the photometric algorithms used may account for PSF variation over the field, to accurately measure the star's fluxes.

\textsc{DAOphot} uses an analytical PSF model (e.g., Moffat), but allows for linear or quadratical variations across the field. Additionally, a look-up table is produced to take the deviations of the true PSFs from the analytical model into account. On the other hand, \textsc{StarFinder} does not use any models, but it does build an empirical PSF by using several stars in the image. When in the presence of PSF variations over the field, the choice of those reference stars is then one critical aspect of \textsc{StarFinder} optimization. The usual way to account for PSF variations with \textsc{StarFinder} is simply by dividing the field into subfields and studying each subfield individually. Of course, there is a trade-off between having enough stars per subfield, so that an accurate PSF model can be extracted, and to keeping the subfield as small as possible to keep the anisoplanatism effect to a minimum. The case of RCW\,41 represents an almost perfect study case for \textsc{StarFinder}, because we have a large number of well-isolated stars uniformly distributed over the field. In that sense, RCW\,41 is not too crowded to face confusion effect (e.g., Schodel et al. \cite{sch10}) or too sparse for many stars to be used to build the PSF model. We experimented with different subfield sizes and found a weak dependence of the results with the number of sub-fields. More precisely, subfields having a size in the range between 30"$\times$30" and 10"$\times$10" were giving the same fluxes results within 2\%. For smaller fields, some subfields were affected by the small number of stars available to build the PSF model ($\leq$1). Larger fields may be  affected by anisoplanatism and could suffer from flux errors of up to 10\%. In the rest of the paper, the local PSF fitting is done on subfields of size 20"$\times$20", shifted by 10" (in x or y) in order to introduce overlap between the frames. Each star will then be measured several times with different set of PSFs, which can be used later as a photometric error estimation.

\begin{table}
\caption{\textsc{StarFinder} parameters used for the star detection and photometry measurement, for each filter. The two values given for the threshold mean that the algorithm does two iterations, with a relative threshold at N sigma each.}            
\label{tabStarFinder} 
\begin{tabular}{ccc}
\hline
\hline
Filter & Threshold & Correlation \\
\hline
$J$ &  [5, 2] & 0.65 \\
$H$ & [5, 3] & 0.7 \\
\ks & [5, 4] & 0.8 \\
\hline
\end{tabular}
\end{table}

The other critical aspect of the \textsc{StarFinder} fine tuning is the choice of the number of iterations, the relative threshold, and the correlation threshold for star extraction at each iteration. 
We adjusted those parameters so that every single faint stars will be detected without too much contamination from the remaining bad or hot pixels or from structures in the background. Our three final reduced images ($J$, $H$, and \ks) have different characteristics in terms of noise and resolution, so this set of parameters has been optimized for each filter. Those parameters are summarized in Table \ref{tabStarFinder}. To detect as many of the faint stars as possible, we found that the best strategy was to set the detection threshold at low levels and then manually inspect each detection in order to remove the spurious ones. 

In the case of RCW\,41, the number of stars is reasonably small ($\sim$ 450 / filter) so that a visual inspection and manual filtering of each detection is possible. Also we previously flagged all the saturated stars, which were defined assuming a minimum detector full well depth of 50000 e-, so that they are excluded from both the star extraction and photometry process. When using AO, saturation may be an issue, especially for images with good  correction. As an example, for the 80s. exposure observations obtained in \ks\ on  January 30, all stars with \ks $<$ 16.5 are saturated in the single-frame image. To tackle this issue, short-exposure images were added to the observational program and are then used to derive the fluxes of the brightest stars. Finally, for four very bright stars present in the field, even the shortest exposure images were saturated, and we used the photometry from the Santos et al. (\cite{san12}) catalog. The final flux catalog contains 549 stars in \ks, 483 in $H$, and 335 in $J$. Over all these detections, 323 stars are commonly detected in the three filters. The 12 stars detected in $J$ but not in $H$ or \ks are either very faint stars, at the detection limit, or stars very close to the image border where noise is higher. Combining the $H$ and \ks catalogs, 475 are commonly detected. There are 8 stars detected in $H$ but not in \ks. Those stars also are on the faint end, or affected by noise or detector features in the \ks filter.

The next step was to convert the measured star fluxes (or instrumental magnitude) into calibrated magnitudes. As for the astrometry calibration, we chose to calibrate our zero-point magnitudes based on the Santos et al. (\cite{san12}) catalog. For that, we used the same set of 100 well-isolated stars. This is done both for the long- and short-exposure images independently. 
A reasonable match is found, with a zero-point uncertainty of $\sim$0.1 magnitude for each filter. This scatter is less accurate than the zero-point uncertainty reported by Santos et al. (\cite{san12}), when they calibrated their {\it{SofI}} images using 2MASS. Ss a sanity check, we also attempt to derive a zero point based on the 2MASS catalog. The mismatch in angular resolution between 2MASS and GSAOI is large, 2MASS resolution being $\sim$1", while we have a factor of 10 better, so only a tenth of well-isolated stars were identified. When cross-calibrating our flux measurement with 2MASS, a better zero-point uncertainty is found, around $0.05$ magnitude for each filter. However, as the number of stars is reduced, statistical uncertainty increases. Furthermore, those stars are on the bright end of the sample, where the photometric error should be more accurate. We also compared the fluxes measured both from \textsc{DAOphot} and \textsc{StarFinder}. 

Overall, very good agreement is found between the two algorithms, and more than 95\% of the stars have instrumental magnitude differences between the two methods of less than 0.05 magnitude. The remaining 5\% of the stars are mainly stars lying close to a bright stars, where the background and wing contamination from neighbors has probably not been done properly. Since \textsc{StarFinder} was designed specifically to deal with AO images, we believe it is more robust to this kind of situation, and in the following we only use the magnitudes derived from \textsc{StarFinder}. As a conservative approach, and also to account for the photometric error found between the \textsc{Iraf} and \textsc{Yorick} data reduction processes, the zero-point accuracy is set to 0.1 mag.

\section{Photometric completeness and contamination}
\label{sec:compcont}

\subsection{Completeness and photometric uncertainties}
\label{ssec:comp}

An important parameter is to derive the completeness limits for each filter. In dense stellar fields, the detectability of a source depends on its flux and on the local stellar density ("crowding"). In the case of RCW\,41, detectability may also be affected by the diffuse background. To determine the completeness of our images, we used simulated stars, embedded in the real images. This method somehow follows the technique described in Gennaro et al. (\cite{gen11}). Stars were simulated based on the \textsc{StarFinder} empirical models: for each subfield location, we used the corresponding local PSF model. This allows to reproduce the real PSFs specificities and variations over the field. Those fake stars are introduced at random positions on each subfields. We used three sets of fake stars, with a total number of respectively 100, 250, and 500 for the whole image. We followed this strategy to make sure that no bias would be introduced by the number of simulated stars, and artificially induced crowding. Finally, for each filter, we simulated a range of magnitude from 13.5 to 22.0, by steps of 0.5 mag. Once simulated stars are added to the images, we ran the \textsc{StarFinder} detection again, using the exact same parameters as the ones used for the analysis of Sect. \ref{ssec:zp}. We can then estimate the number of simulated stars that are properly detected and compute a completeness limit. At the same time, we can also estimate the photometric error by computing the standard deviation of the difference between the measured and simulated magnitudes. 

\begin{figure*}
   \centering
  \begin{tabular}{cc}
  \includegraphics[width= 0.4\linewidth]{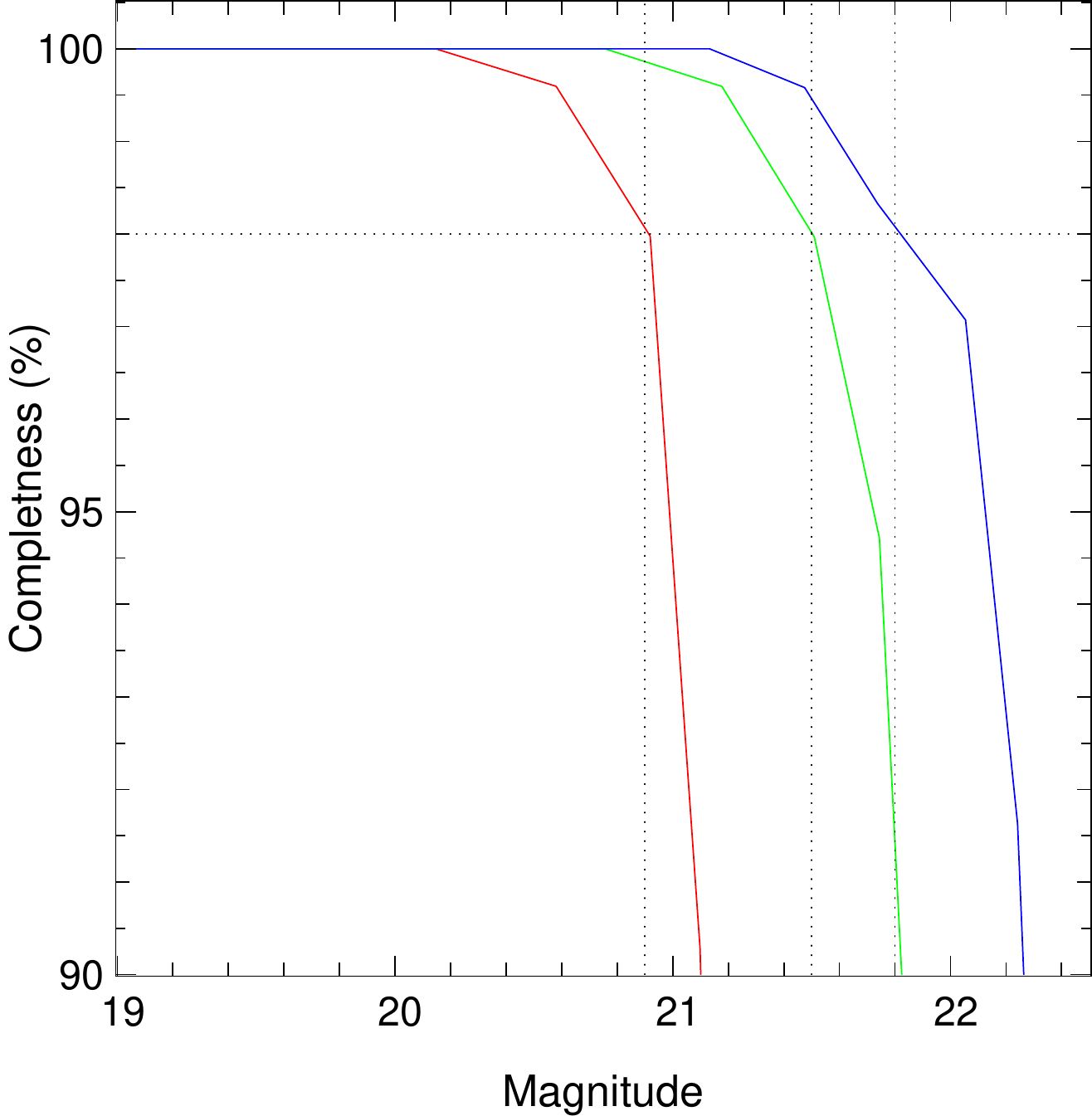} &
  \includegraphics[width=0.4\linewidth]{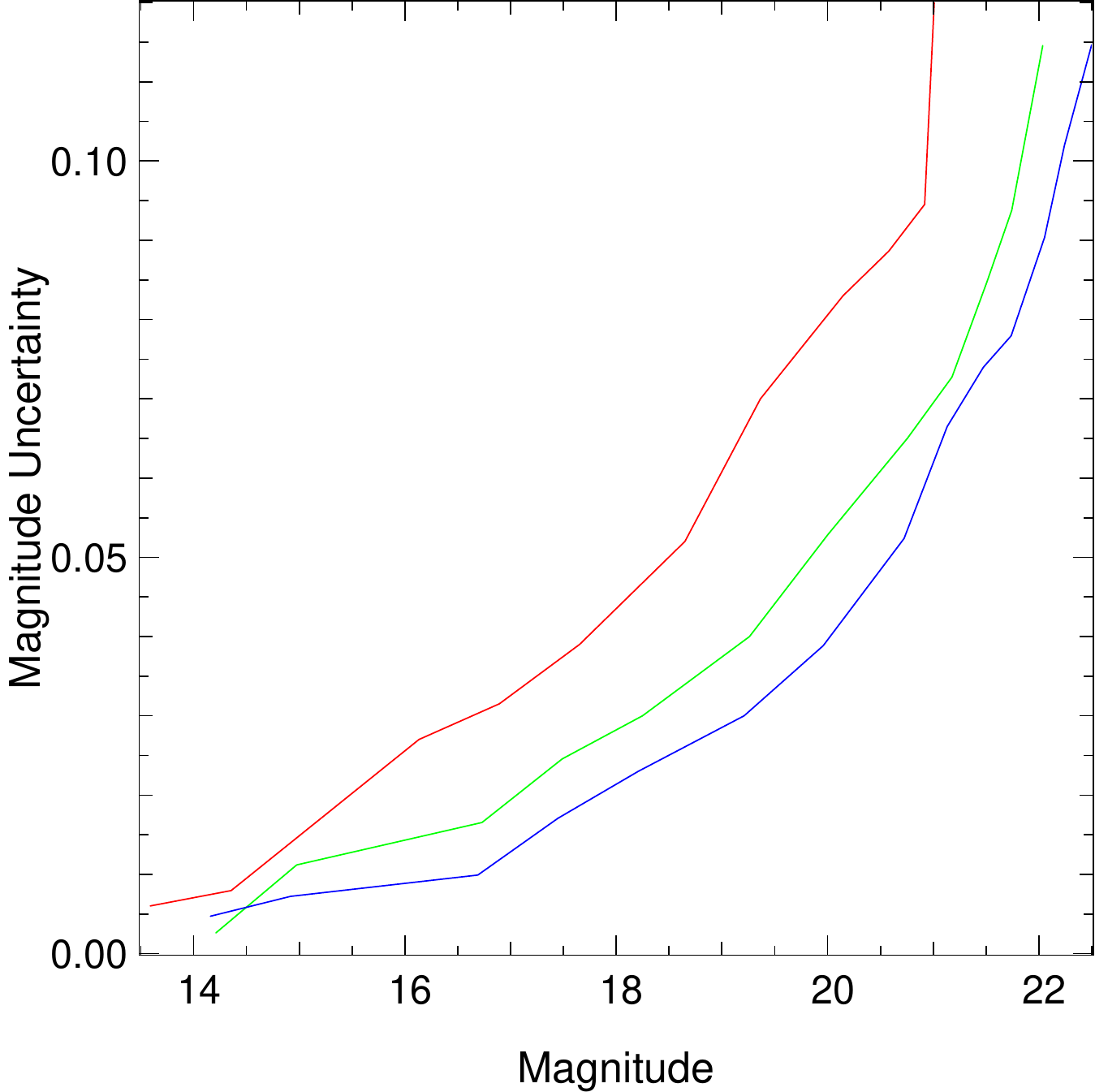}  
  \end{tabular}
   \caption{Left: Completeness limit estimated from simulations. Right: photometric error estimated from simulations. Red is \ks, green is $H$, and $J$ is blue.}
              \label{fig:completude}
\end{figure*}

Results for the completeness limit are presented in Fig. \ref{fig:completude} (left). The values obtained, assuming 98\% of completeness, are ($J$,$H$,\ks)$_{limit}$ = (21.8, 21.5, 20.9). This is (3.2, 3.1, 3.1) magnitudes deeper than the limiting magnitude derived by Santos et al. (\cite{san12}). A completeness level of 98\%, such as the one chosen in the study, is quite conservative compared to similar studies found in the literature. We used such a conservative number to avoid any bias towards the faint population, which is the population of main interest in this cluster. For reference, the detection limits are ($J$,$H$,\ks)$_{detect}$ = (23.0, 22.5, 21.9). 

Based on the completeness limits, one can derive the mass detection limit, following for instance the Baraffe et al. (\cite{bar03}) isochrones. Assuming the 1 Myr isochrones, an extinction of $A_k$ = 1.0 (see Santos et al. \cite{san12} and Sect. \ref{ssec:conta}), and a distance of 1.3 kpc (Roman-Lopes et al. \cite{rom09}), we derive that the apparent magnitude of a M = 0.006$M_{\sun}$ star is \ks = 21, which does correspond to our completeness limit. We can then presumably detect stars well below the hydrogen-burning limit of 0.08 $M_{\sun}$, and potentially prove the presence of brown dwarf candidates. Pushing the cluster age to 5 Myr would raise the mass detection limit to 0.015 $M_{\sun}$, which is still well below the hydrogen-burning limit. We can confidently state that the completeness limit of our observations probes the brown dwarf regime for any plausible age of the cluster.

Results for the photometric uncertainties are presented in Fig. \ref{fig:completude} (right). Those errors must be quadratically combined with the 0.1 mag. zero-point error derived above. This gives the final photometric accuracy. No bias was found between the 100, 250, and 500 simulated stars cases. This confirms that crowding does not have a major impact on the photometric accuracy for RCW\,41.

\subsection{Magnitude histograms}

\begin{figure*}
   \centering
  \begin{tabular}{ccc}
  \includegraphics[width= 0.3\linewidth]{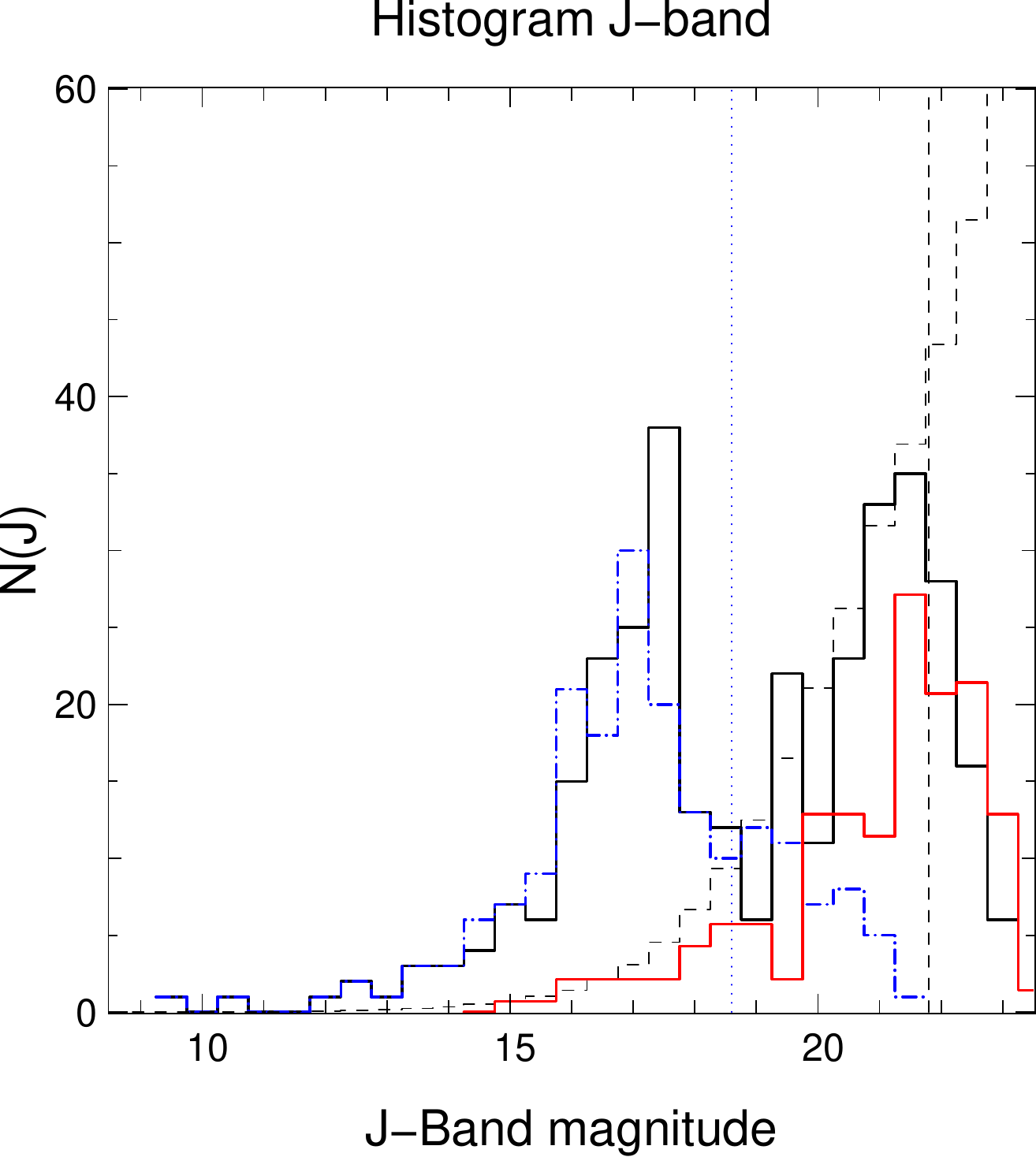}  &
  \includegraphics[width=0.3\linewidth]{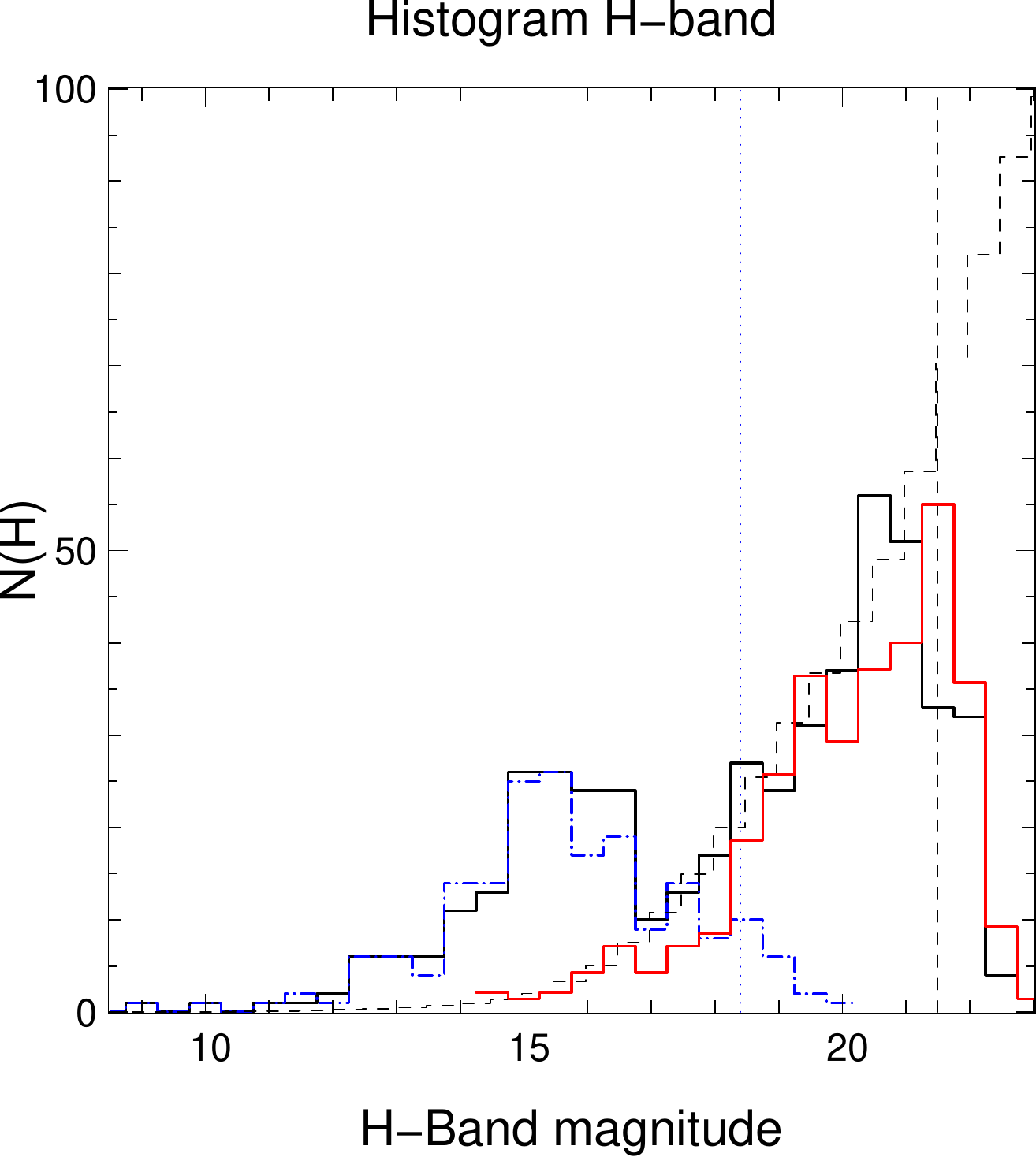}  &
  \includegraphics[width=0.3\linewidth]{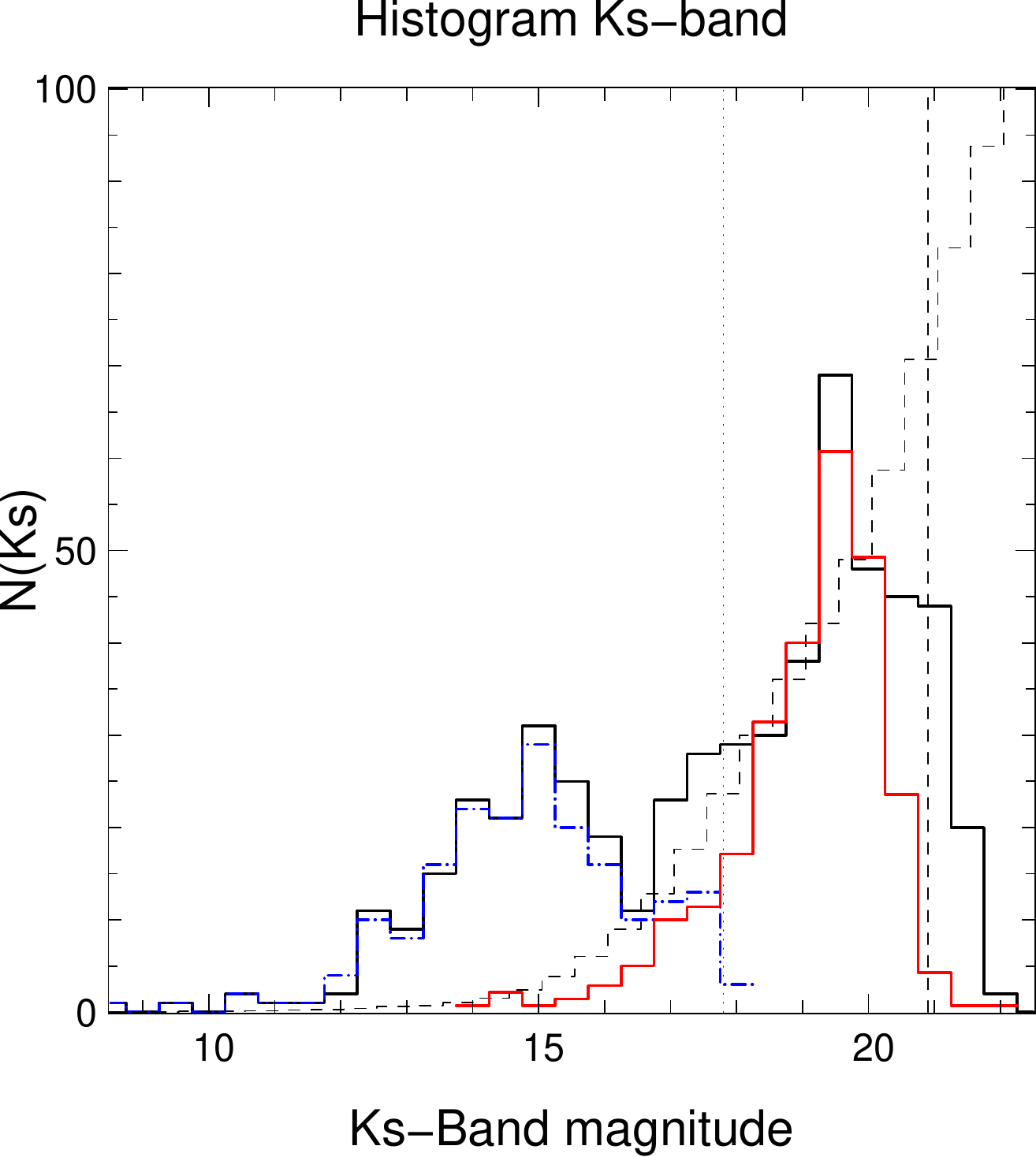}   
  \end{tabular}
   \caption{Histogram of the stars for the three filters. GeMS/GSAOI data are the black histograms. Blue histograms are taken from the Santos et al. (\cite{san12}) catalog, only for number of stars per magnitude bin for the same region as the one covered by GSAOI. For each data set, the completion limit has been added. Dashed histogram show star counts from the Besancon model. Red histogram are the star counts from the control field.}
              \label{fig:histo}
\end{figure*}

\begin{figure*}
   \centering
  \begin{tabular}{cccc}
  \includegraphics[width= 0.2\linewidth]{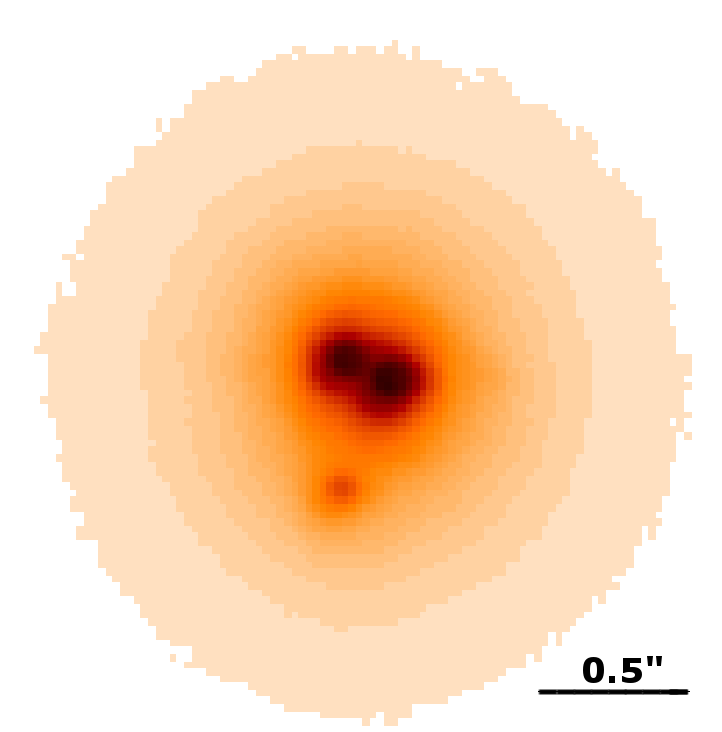}  &
  \includegraphics[width=0.2\linewidth]{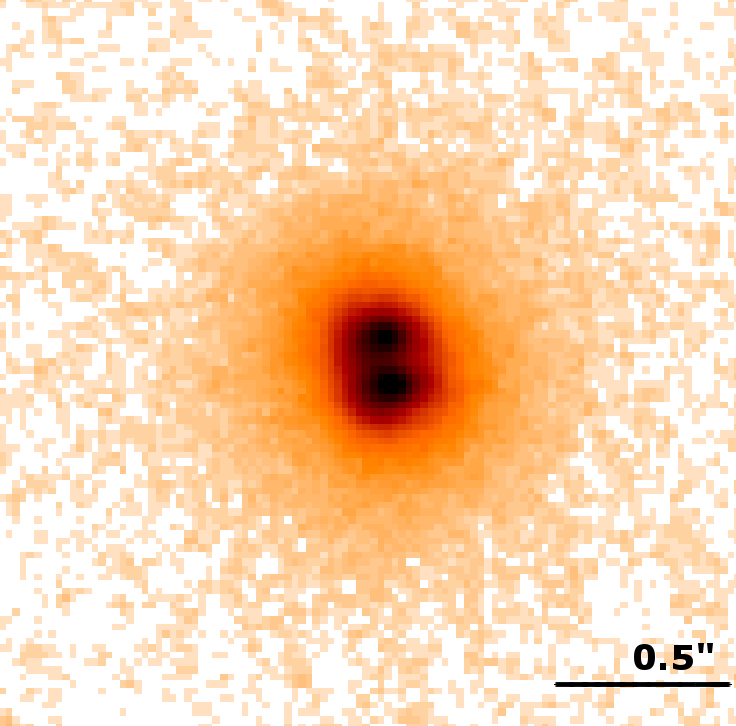}  &
    \includegraphics[width=0.2\linewidth]{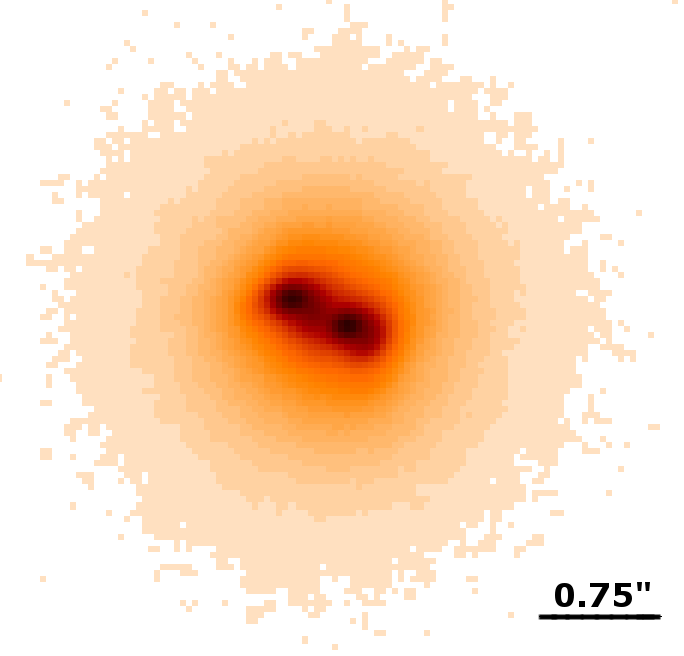}  &
    \includegraphics[width=0.2\linewidth]{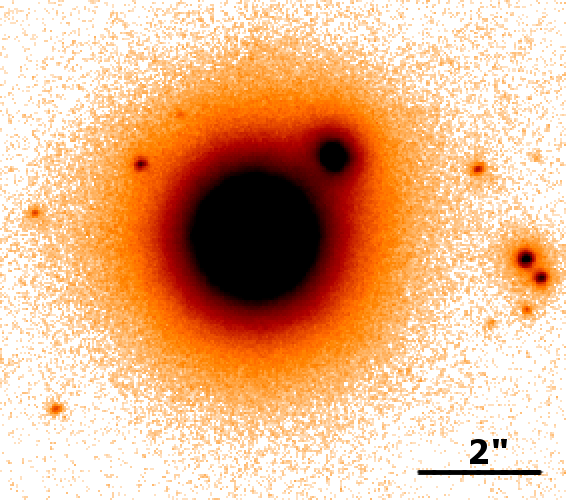}   \\
      \includegraphics[width= 0.2\linewidth]{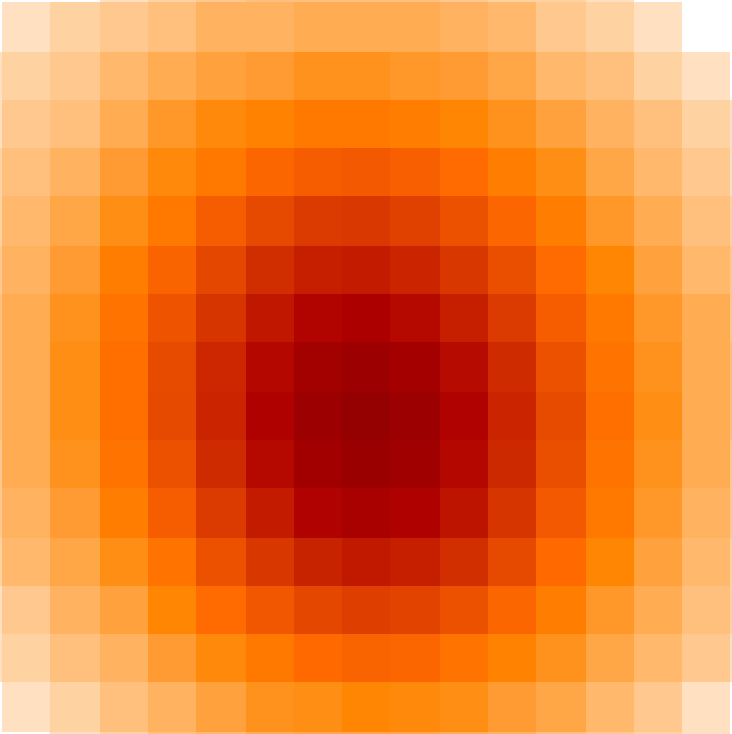}  &
  \includegraphics[width=0.2\linewidth]{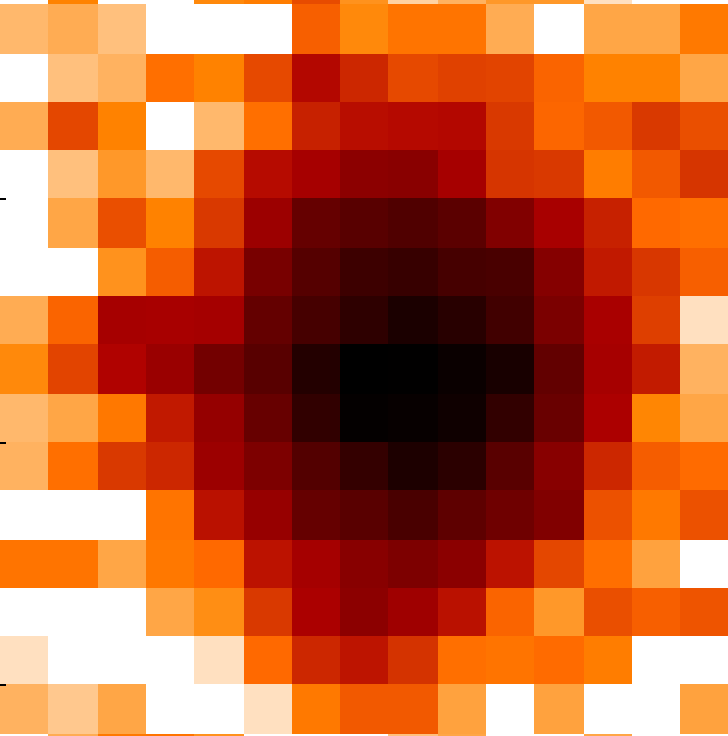}  &
  \includegraphics[width=0.2\linewidth]{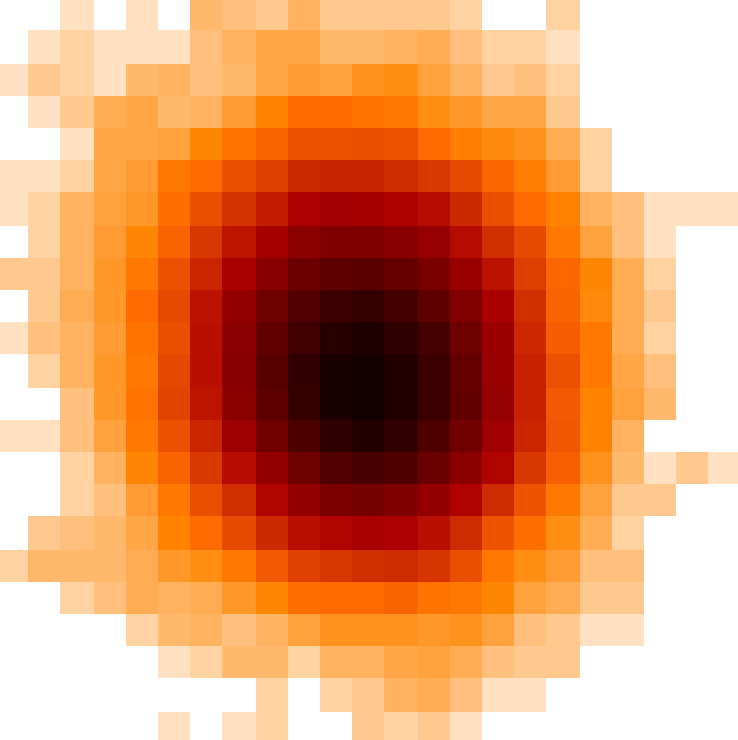}   &
    \includegraphics[width=0.2\linewidth]{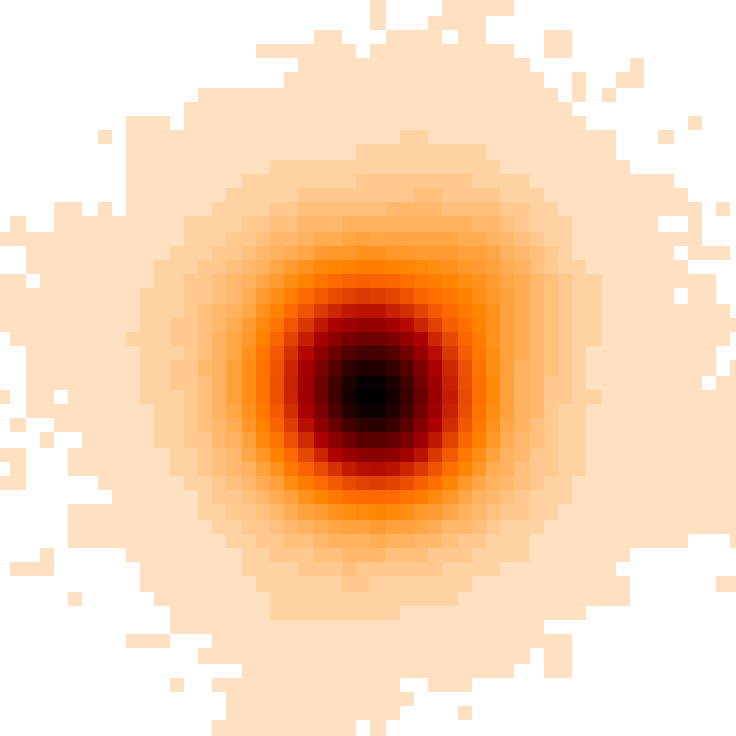}  
  \end{tabular}
   \caption{Example of multiple systems taken from the $H$-band, resolved by GeMS/GSAOI (top row) and un-resolved by {\it{SofI}} (bottom row). The separation of the main components is, from left to right, [120, 140, 280, 1600] mas ([156, 182, 364, 2080] AU, respectively). The $H$-band magnitude measured on the first three GeMS/GSAOI images are, from left to right [13.5, 14.2, 16.6], [18.4, 18.6], and [15.8, 15.9] for the multiple systems, while the magnitude derived from the {\it{SofI}} images are 13.0, 17.6, and 15.1, respectively.}
              \label{fig:PSFs}
\end{figure*}

\begin{figure}
   \centering
  \begin{tabular}{c}
  \includegraphics[width= 0.9\linewidth]{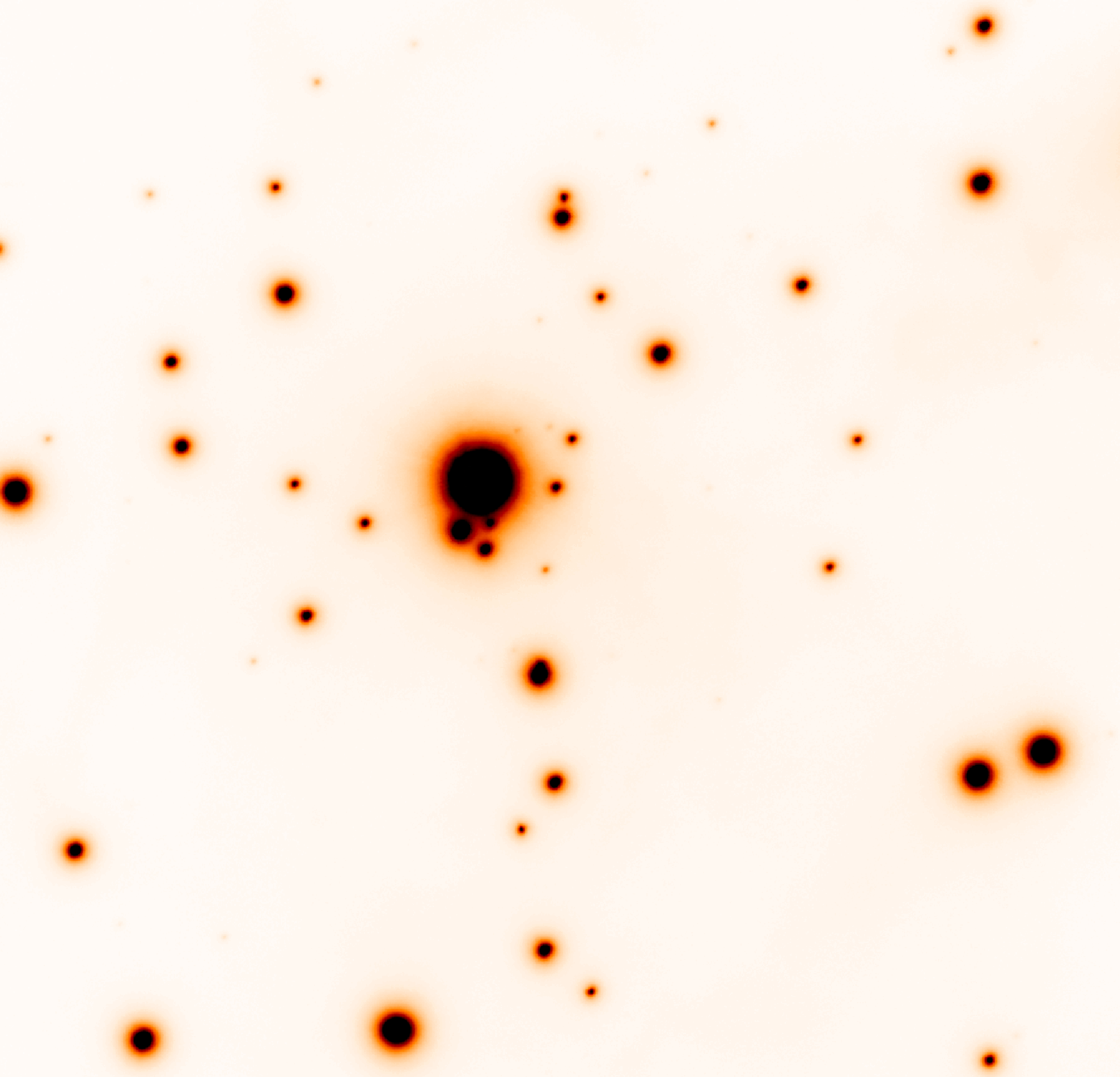} \\
    \includegraphics[width=0.9\linewidth]{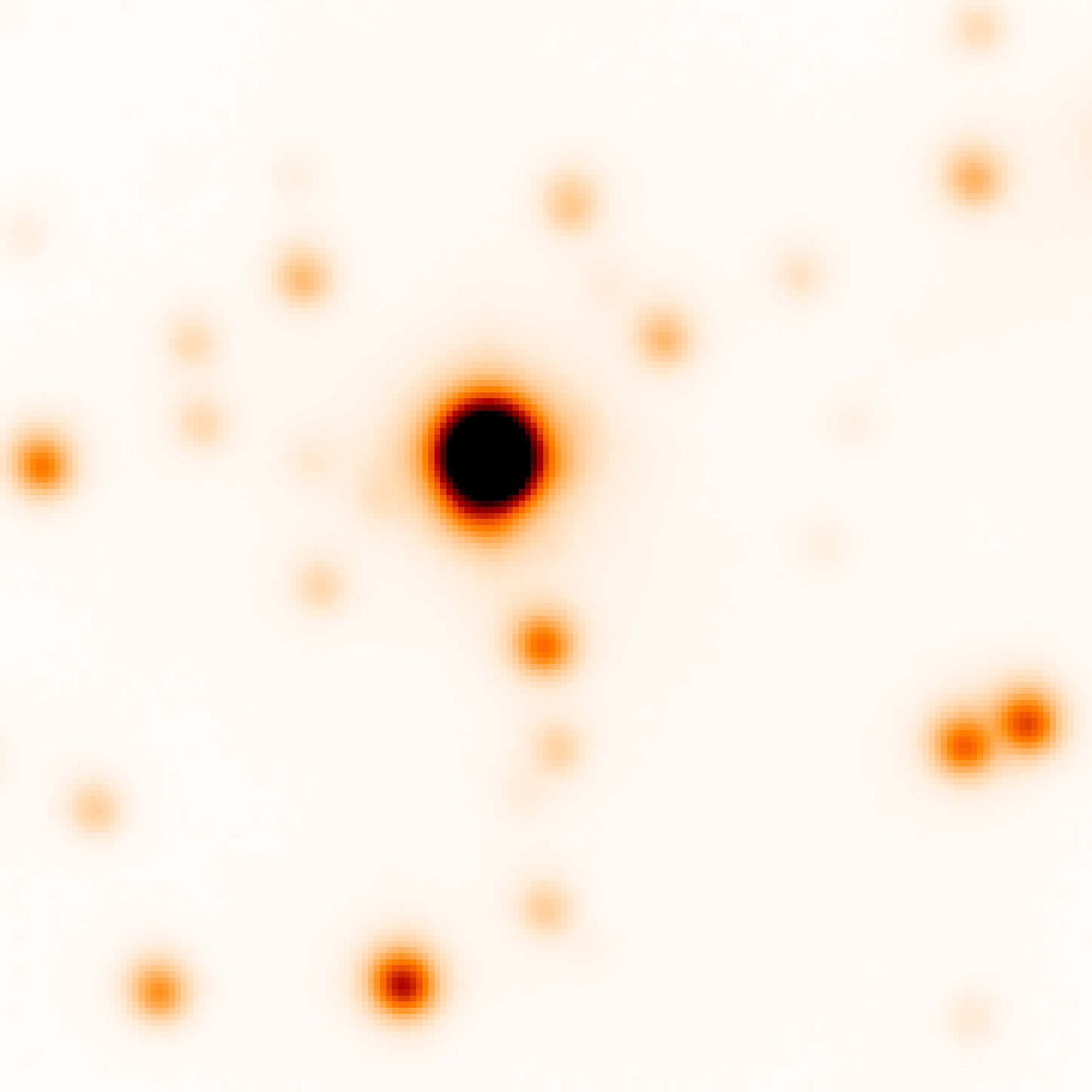}  
  \end{tabular}
   \caption{Close view of the center part of the main cluster around Obj1a (RA = 09$^h$16$^m$43.47$^s$ ; DEC = -47\degr56 \arcmin 22.86\arcsec), by GeMS/GSAOI (top row), and by {\it{SofI}} (bottom row) taken from the $H$ band. North is up and east is right, and the FoV is 25\arcsec$\times$25\arcsec. The same stretch (cutting pixels at 1\% from the minimum and 99\% from the maximum) and linear scale has been used for the two images.}
              \label{fig:PSFs2}
\end{figure}

In Fig. \ref{fig:histo}, we show the star histograms for the three GSAOI filters used. For reference, histograms derived from the Santos et al. (\cite{san12}) catalog are overplotted, but only for the stars covering the same region of the cluster. For each data set, the completion limit derived in Sect. \ref{ssec:comp} is added. \\
Looking at Fig. \ref{fig:histo}, the first striking element is that the GeMS/GSAOI data shows, for each filter, a secondary distribution of star at faint magnitudes, which could not have been detected by the previous studies owing to their limiting magnitude. The nature of this second peak is discussed in Sect. \ref{ssec:conta}. We also note that, even for the common region of the histogram, some discrepancies between the GeMS/GSAOI catalog and the Santos et al. (\cite{san12}) exist. We looked in more detail into these differences and found that all those stars were actually multiple systems that were not spatially resolved by the previous {\it{SofI}} images. This is also shown by the fact that, for the common magnitude range, the GeMS/GSAOI histogram detects fainter stars, because those stars are actually resolved in fainter systems. As an example, Fig. \ref{fig:PSFs} shows some representative multiple stars, which are taken from the $H$ band and properly resolved by GeMS/GSAOI and the corresponding un-resolved {\it{SofI}} images. Figure \ref{fig:PSFs2} shows a close-up view of the $H$ band central part of the main cluster, illustrating the gain in spatial resolution and limiting magnitude.

\subsection{Contamination}
\label{ssec:conta}
Foreground and background stars are contaminants for studies of star clusters, hampering the stellar distributions of interest. Young cluster environments are rich with dust, potentially reducing the density of background stellar contamination. However, in the case of deep observations close to the galactic plane, such as the one presented here, background contamination may become significant. 
To estimate the foreground and background contamination, we first used star counts from the control field, corrected for the surface difference between the control and the science fields. Those star counts are shown in Fig. \ref{fig:histo}. The control field observations were done without MCAO corrections, so the resulting resolution and completeness limit are lower than those of the science fields. The strategy we followed was to then compare the control field star counts with star counts derived from the Besancon galaxy model (Robin et al. \cite{rob03}) and use the model star counts for the faint magnitudes. According to the Besancon model, foreground stars counts are negligible, and we only considered background stars. Background stars taken either from the control field or from the Besancon model must be corrected for the extinction in the light of sight of the cluster. We estimated the extinction following the same method as in Gutermuth et al. (\cite{gut05}), by deriving the two-dimensional extinction maps ($A_k$ maps) using the $H$-\ks colors of all stars detected in both filters. 
The extinction shows a clear gradient, with higher extinction values on the northwestern part of the field, the mean cluster extinction value being $A_K^{\mbox{cluster}}$ = 1.0. For comparison, we computed the same extinction index for the control field. The averaged $A_K$ is 0.2, and it is uniformly distributed over the control field. The star counts derived from the Besancon model are then shifted toward redder magnitudes, based on the average extinction difference ($\delta A_K$) between the cluster and the control field: $\delta A_K$ = $A_K^{\mbox{cluster}}$ - $A_K^{\mbox{field}}$ = 0.8. 

Similar analysis was performed for $J$ and $H$. The extinction law, for the transformation between \ks and other NIR wavelength is done following Rieke \& Lebofsky (\cite{rie85}). Star counts from the control field, corrected for the surface difference between control and science field, are shown in Fig. \ref{fig:histo}, as is the star count from the Besancon model. For $H$ and \ks filters, both the control field and model agree and tend to demonstrate that star field contamination is fully consistent with the secondary peak of faint stars. For the $J$-band filter, a discrepancy between the field star and the model count appears around 18<$J$<20, and no clear explanation for this difference has been found. The $J$-band star counts will then be excluded when deriving the cluster IMF (Sect. \ref{ssec:imf}). 
%

\begin{figure}
   \centering
  \includegraphics[width= 1.\linewidth]{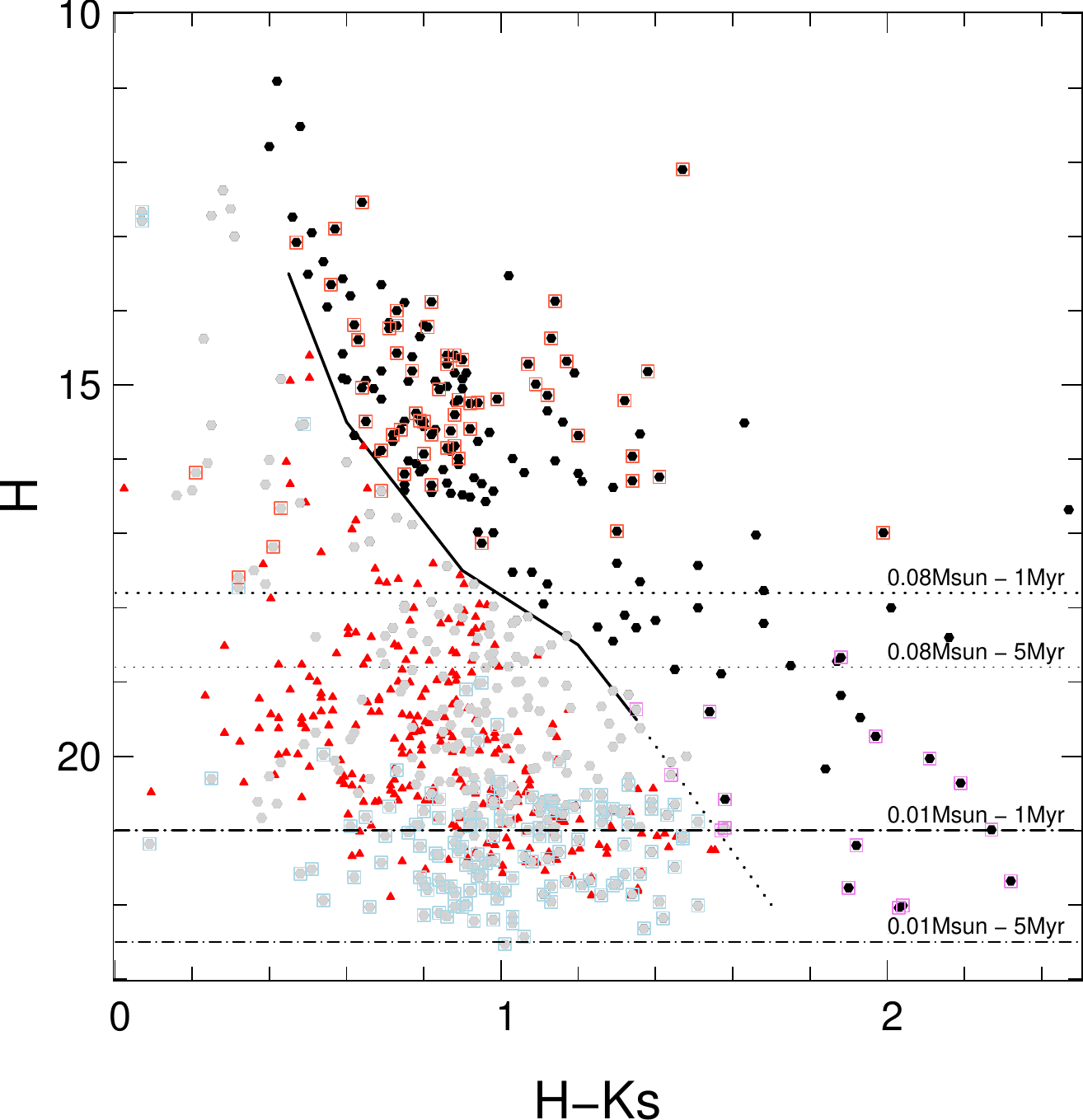} \\
  \caption{($H$ $\times$ [$H$-\ks]) CM diagram for all stars detected by GeMS/GSAOI. The solid line shows the statistical separation between cluster members (black points) and contaminating field stars (gray dots), as derived from Santos et al. (\cite{san12}). The dotted line is an extrapolation of the solid line at faint magnitudes. Red triangles are stars from the control field. Stars flagged with red open squares are potential NIR excess stars (see Sect. \ref{ssec:yso}). Stars flagged with violet open squares are stars detected in both $H$ and \ks, but not in $J$. Stars flagged with blue open squares are also detected only in $H$ and \ks, but they are likely to be field stars. }
              \label{fig:cmdcolorcut}
\end{figure}

A different approach to identifying a cluster member is to directly differentiate likely noncluster members on the basis of their location in a CM diagram. Santos et al. (\cite{san12}) performed a statistical analysis of the cluster star and control field star distribution in the ($H$ $\times$ [$H$-\ks]) and (\ks $\times$ [$J$-\ks]) CM diagram, to draw a separation criterion between cluster members and contaminants. A somewhat similar approach is followed in Harayama et al. (\cite{har08}), where they use the ($J$ $\times$ [$J$-\ks]) CMD to perform a \textit{\emph{color cut}} to segregate cluster members alone. If we apply the color-cut defined in Santos et al. \cite{san12}, we find the results presented in Fig. \ref{fig:cmdcolorcut}. Stars located on the lefthand part of the solid line are assumed to be field stars, not part of the cluster. Stars from the control field are overplotted and are all be detected as field stars. Field stars are mainly faint stars, which is consistent with what was found in Fig. \ref{fig:histo}. Finally, we also looked at the spatial distribution of the stars located on the secondary peak. As expected for field stars, those stars are uniformly distributed over the field, without a clear correlation with the main cluster structures. Quantitative comparison of the contamination correction methods is presented in Sect. \ref{ssec:klf}.

\section{Color diagrams and young stellar object contents}
\label{sec:cd}

\subsection{Color-color diagram}
\label{ssec:ccd}

\begin{figure}
\centering
  \includegraphics[width= 1.\linewidth]{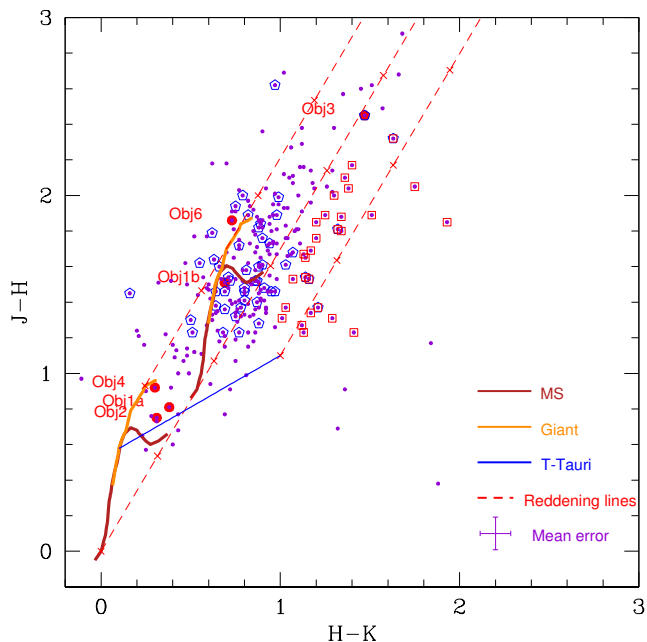} \\
\caption{
([$J-H$] $\times$ [$H-Ks$]) CC diagram for all the sources detected in our GeMS FOV. The  solid (fire brick) and thick dashed (dark orange) curves represent the unreddened (bottom) and reddened (top) MS and giant branches (Bessell $\&$ Brett \cite{bes88}), respectively. The blue solid line is the locus of intrinsic T-Tauri stars (Meyer et al. \cite{mey97}). The colors of the curves and the T-Tauri line are converted to the 2MASS system using the relation given in Carpenter et al. (\cite{car01}). The parallel dashed lines are the reddening vectors drawn from the tip of the giant branch (``left reddening line''), from the base of the MS branch
(``middle reddening line''), and form the TTS  branch (``right reddening line''). The known O or early-B stars are shown as solid dots. The NIR-excess sources identified from  ([$J$-$H$] $\times$ [$J$-[3.6]]) and 
([$H$-\ks] $\times$ [\ks-[4.5]]) CC diagrams are marked as polygons (see Fig. \ref{fig:irac}). The NIR-excess sources identified from this diagram are marked as squares.
The error bars in the bottom right corner show average errors in the colors.}
\label{fig:cmd}
\end{figure}

The CC diagram is shown in Fig. \ref{fig:cmd}. Plotted are the 259 sources detected commonly in \jhk, for which the photometric error is lower than 0.15 mag in each band. 
In this diagram, most of the sources are distributed in the reddening band of the main sequence, which could be a mixture of reddened field stars and sources having warm circumstellar dust, characteristic of young pre-main sequence (PMS) objects (Lada \& Adams \cite{lad92}). Stars lying below the middle reddening vector (i.e., below the MS reddening line) are expected to be associated with NIRexcess from circumstellar disk. We can then use the star location on the CC diagram to identify the NIR-excess objects. However, since observations of star forming regions are likely to be affected by high levels of extinction, it would make dusty objects along the line of sight indistinguishable from NIR excess sources. 

As an attempt to disentangle such sources, we only selected stars whose excess is more than 1$\sigma$ (where $\sigma$ is the mean color error) from the MS reddening line and with a ($J$-$H$) color greater than 1.2 mag. This color cut is chosen based both on the distribution of the control field sources, which mostly lie below a ($J$-$H$) color of 1.2 mag (see also Fig. 5 in Santos et al. \cite{san12}), and on the IRAC-identified NIR-excess sources (see Sect. \ref{ssec:yso}), which are also found to be located above a ($J$-$H$) color of 1.2 mag. With these selection criterion, we may miss a few massive sources, but we expect to limit the degree of contamination in the final NIR-excess sources catalog. The 28 selected NIR-excess sources are marked in the CC diagram.

\subsection{IRAC identification of young stellar objects}
\label{ssec:yso}

The circumstellar emission from the disk and envelope in the case of YSOs dominates at long wavelengths, where the spectral energy distribution (SED) significantly deviates from the pure photospheric emission. It is then preferable to use the longer wavelengths available for identifying infrared excess stars. For example, Haisch et al. (\cite{hai00}) found a fraction of NIR-excess sources of $\sim$80\% for NGC\,2024 with $JHKL$ observations, while a percentage of 54\% is found when only $JHK$ bands are used.
 
\begin{figure}[!ht]
\centering
\resizebox{9.0cm}{9.0cm}{\includegraphics{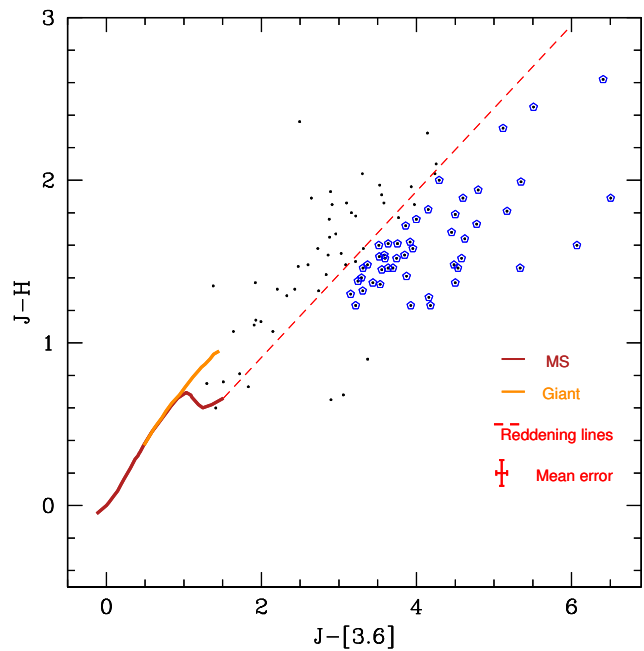}}
\resizebox{9.0cm}{9.0cm}{\includegraphics{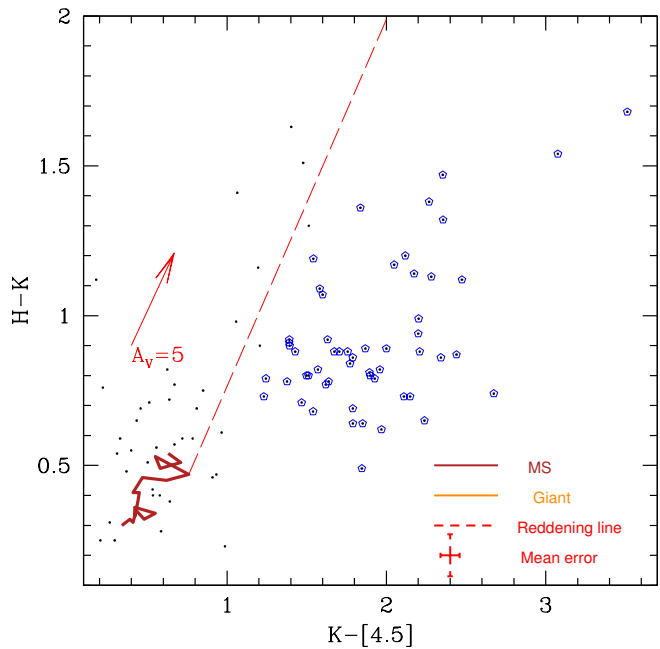}}
\caption{
{\it Top:}
([$J$-$H$] $\times$ [$J$-[3.6]]) CC diagram of all the common sources detected in our GeMS FOV. The curved solid line (firebrick) is the MS locus. The long dashed line (red) 
represents the reddening vector from the tip of an M6 dwarf. The IR-excess sources identified from this diagram are marked by open polygons (see text for details).
{\it Bottom:}
The ([$H$-\ks] $\times$ [\ks-[4.5]]) CC diagram. The curved solid line (firebrick) is the MS locus of late M-type dwarfs. The long dashed line (red) represents the reddening vector from the tip of an M6 dwarf. The NIR-excess sources identified from this diagram are marked by polygons (see text for details).
The error bars in the bottom right corner of both diagrams show the average errors in the colors.
}
\label{fig:irac}
\end{figure}

 In this work, we used IRAC observations in combination with our \jhk observations to reliably identify more NIR excess sources than those already found in Sect. \ref{ssec:ccd}. To do so, we matched the 3.6 and  4.5 $\mu$m  catalogs with the deep NIR catalogs using a 1$\farcs$2 radial matching tolerance, which corresponds to one pixel of the IRAC images. We then used the ([$H$-\ks] $\times$ [\ks-[4.5]]) and ([$J$-$H$] $\times$ [$J$-[3.6]]) color diagrams to identify NIR excess sources (see Fig. \ref{fig:irac}). In these diagrams, the sources located to the right of the MS reddening vector are likely to be YSOs with NIR excess. Following a similar approach to the one discussed in Sect. \ref{ssec:ccd}, NIR-excess candidates are selected as sources whose excess is more than 1$\sigma$ (where $\sigma$ is the color error) from the MS reddening line. \\
With the above approach, we identified 57 candidate YSOs with NIR-excess emission, five of them in common with \jhk ones found in the previous section. This number is probably a lower limit, since many sources falling in the bright nebulous regions were not detected in the 3.6 and 4.5 $\mu$m  bands.\\ 
Combining the candidate YSOs identified in Sect. \ref{ssec:ccd} (28 sources) with the ones identified based on IRAC photometry (52 new sources), we end up with a total of 80 NIR-excess sources for the area covered by our \jhk observations.   

\subsection{Extinction}
\label{ssec:exc}

A first estimation of the extinction has been done in Sect. \ref{ssec:conta}, based on the method followed by Gutermuth et al. (\cite{gut05}). In this section, we want to consolidate the derived extinction value by estimating it directly from the CC diagram. To derive the extinction of the region, we selected only the sources located within the MS reddening band of the ([$H$ - \ks] $\times$ [$J$ - $H$]) diagram in order to avoid the contribution from giant stars along the line of sight and the NIR-excess sources of the region. We then derived the extinction values of these selected sources by tracing them back along the reddening vector to the TTS locus, assuming that most of the sources are PMS in nature. We used the extinction laws of Rieke \& Lebofsky (\cite{rie85}), in which \av= E($J$ - $H$)/0.107 and \av= E($H$ - $K$)/0.06. The mean extinction value for the sources above the TTS locus was found to be \av = 8.5 $\pm$ 2.2 mag. This value is consistent with the mean extinction value derived for the region from spectroscopic observations (Roman-Lopes et al. \cite{rom09}). 

According to the survey by Roman-Lopes et al. (\cite{rom09}), stars labeled as Obj1a, Obj2, and Obj4 in the ([$H$ - \ks] $\times$ [$J$ - $H$]) diagram belong to the most massive objects of the region with spectral type B0V, O9V, and B7V, respectively. The O-type and early B-type MS stars shows a narrow range in their intrinsic colors. Adopting intrinsic ($J$-$H$) colors of OB stars from Koorneef et al. (\cite{kor83}) and using the measured ($J$-$H$) photometric colors, we derived a visible extinction in the range of 7.2 to 8.8 mag with a mean $\sim$ 8.6 mag. This is in very good agreement with the estimation based on the CC diagram. This value is also perfectly consistent with the one derived in Sect. \ref{ssec:conta}, of $A_K$ = 1.0 $\pm$ 0.2 mag, considering $A_K$/\av = 0.112 (Rieke \& Lebofsky \cite{rie85}). For the further analyses, we use \av $\sim$ 8.5  mag as the average visual extinction of the region.

\subsection{Age of RCW\,41}
\label{ssec:age}

The ages of young clusters are typically derived from post–main-sequence evolutionary tracks for the earliest members if significant evolution has occurred or by fitting the low-mass contracting population with theoretical PMS isochrones. Based on an analysis of the ($H$ $\times$ [$H$-\ks]) and (\ks $\times$ [$J$-\ks]) diagrams, Santos et al. (\cite{san12}) derived an age in the range 2.5 to 5 Myr for the region. Lada \& Lada (\cite{lad03}) note that the embedded phase of star cluster evolution lasts 2 to 3 Myrs, and clusters older than 5 Myrs are rarely associated with molecular gas. In the case of RCW\,41, the presence of molecular gas has been confirmed (Pirogov et al. \cite{pir07}) In addition, the detection of maser sources is an indication of early stages of star formation in a dense circumstellar environment. These observations suggest that the cluster is young, at least younger than 5 Myr.

\begin{figure}
\centering
\resizebox{10.0cm}{10.0cm}{\includegraphics{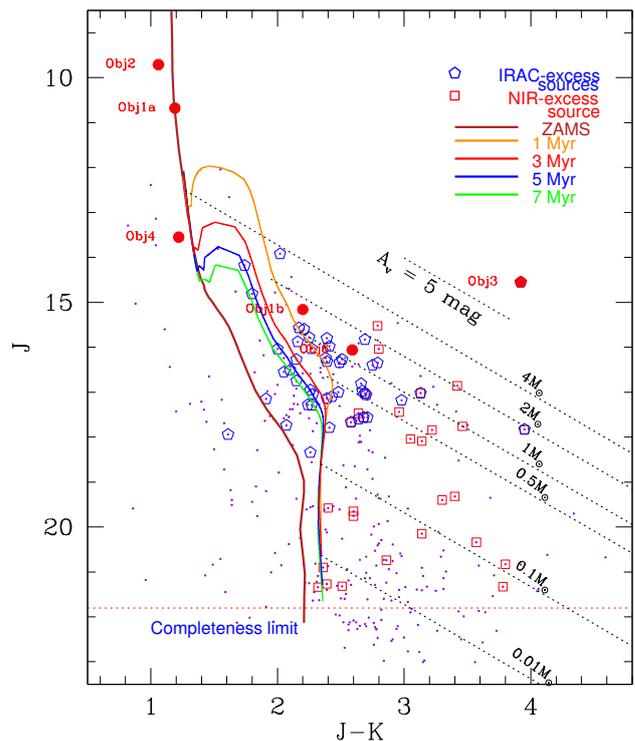}}
\caption{ ($J$ $\times$ [$J$-\ks]) color–magnitude diagram of the stars in the direction of the RCW\,41 region.   
Small dots show all the stars observed in the target field. 
The squares represent the NIR-excess sources identified on the basis of the NIR CC diagram, whereas the polygon are the sources that are using NIR plus IRAC bands. 
The  thin solid lines are the PMS isochrones of Pisa models (see Bell et al. \cite{bel14} for a detailed description) for stars of solar metallicity and ages of 1, 3, 5, and 7 Myr from right to 
left, corrected for a distance 1.3 kpc and average reddening \av = 8.5 mag. 
The reddening vectors corresponding to the 0.01, 0.1, 0.2, 0.5, 1, 2, and 4 $\msun$ of the 1 Myr isochrone are drawn in dotted slanting lines. 
}
\label{fig:jjmk}
\end{figure}

To ascertain the age of the stellar members, we attempted a quantitative age determination with the help of the near-infrared CM diagrams. We based our age estimation on the ($J$ $\times$ [$J$-\ks]) CMD shown in Fig. \ref{fig:jjmk}. The zero-age main sequence (ZAMS) of Marigo et al. (\cite{mar08}) and PMS isochrones from the Pisa models (Bell et al. \cite{bel14}) are also shown. These PMS models are based on an interior model of Tognelli et al. (\cite{tog11}, \cite{tog12}) with the Allard et al. (\cite{all11}), Brott \& Hauschildt (\cite{bro05}), and Castelli \& Kurucz (\cite{cas03}) bolometric corrections and additional empirical corrections as discussed in Bell et al. (\cite{bel14}). 
The isochrones were corrected for a distance of 1.3 kpc and a visual extinction of 8.5 mag (see Sect. \ref{ssec:exc}). 
From this CMD, it is quite apparent that our $J$-band detection limit is deep enough to detect a  0.01$~$\msun$~$star at and age of 1 Myr and \av = 8.5 mag. From Fig. \ref{fig:jjmk}, one can notice a vertical distribution of stars, parallel to the reddened MS shown, but shifted toward bluer magnitudes. This distribution is probably due to foreground stars, indicating that stars belonging to RCW\,41 or farther away are on its right side. 

\begin{figure*}
   \centering
  \includegraphics[width= 0.95\linewidth]{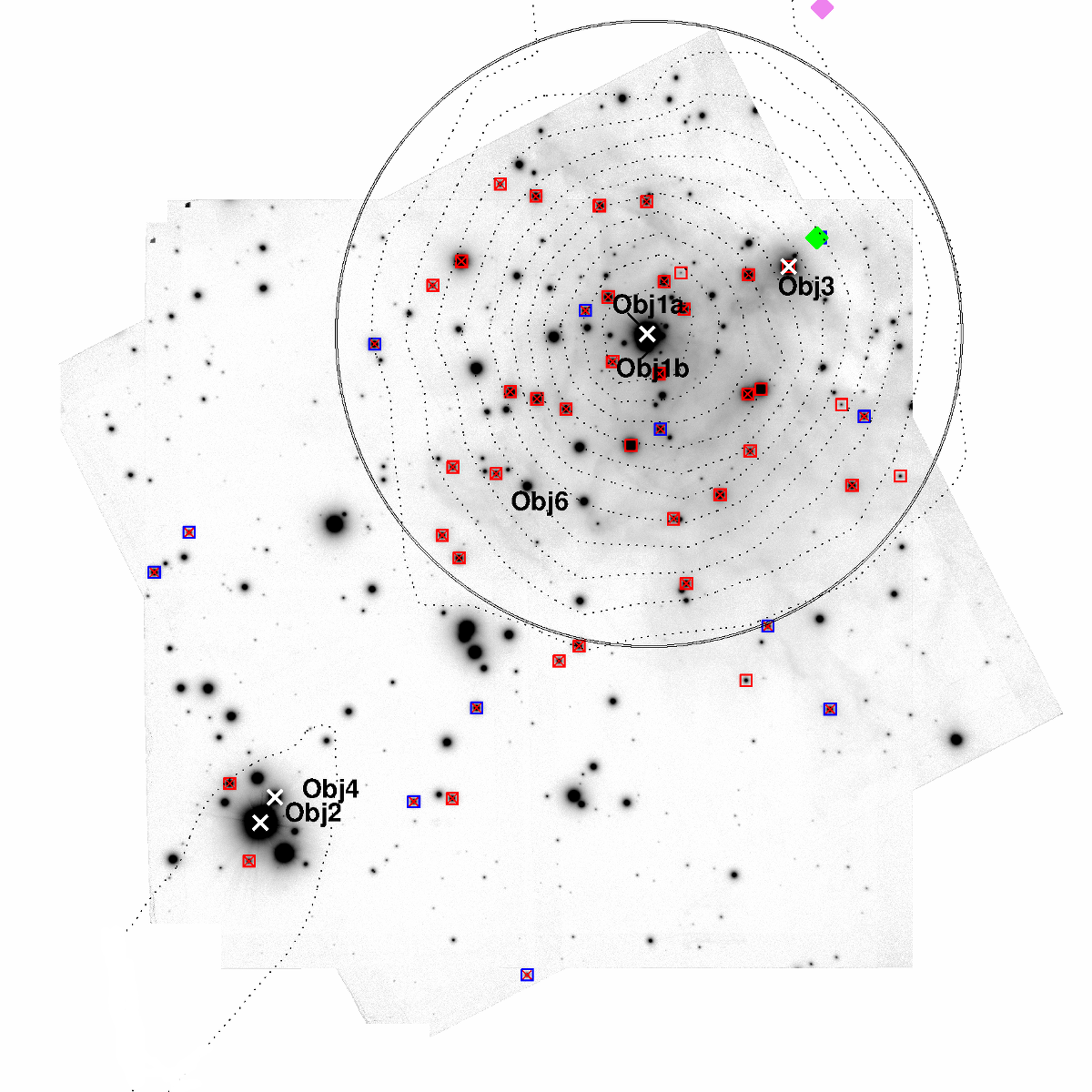}
  \caption{Spatial location of the YSOs identified with {\it Spitzer}. Sources with  red crosses are those with an age estimate. Sources with a red square are the YSOs with an age $<$ 3 Myr, and blue are the YSO with an age $>$ 3 Myr. The green diamond is a Methanol maser (Caswell et al. \cite{cas10}), and the violet diamond (off the GSAOI fields) is a water maser (Breen et al. \cite{bree11}). Other masers have been detected outside the GSAOI field in the northwest part of the cluster. White crosses show the massive stars spectroscopically identified by Roman-Lopez et al. (\cite{rom09}). Dotted lines show MSX isophotes at 14.7 $\mu$m (band D).The full-line circle at the edge 14.7 $\mu$m emission identify the contour of the young main cluster region.}
              \label{fig:nirex}
\end{figure*}

In the absence of proper motion studies or spectroscopic information, the distinction of true low-mass members from the contaminating field stars projected along the line of sight is difficult. However, since we have identified a significant number of likely NIR-excess candidates based on IRAC and NIR photometry, we therefore used these NIR-excess sources to derive an approximate age of the cluster. 
Owing to the lower sensitivity of the IRAC observations, and combined with the fact that low-mass stars do not emit significant excess emission at \jhk bands, our NIR-excess sample is biased toward the massive ends. 
But although the number of our NIR-excess sources is small, it should nonetheless reflect the age of the region. \\
To determine the age of the cluster, we only used the IRAC identified NIR-excess sources, because most of them are located on the MS reddening line of ([$J$-$H$] $\times$ [$H$-\ks]) diagram, suggesting they possess little or no circumstellar emission at these bands. This should limit the impact of NIR excess flux in the age determination. 
If we focus on those sources, one can see that their distribution in Fig. \ref{fig:jjmk} does not fall along a single isochrone, but instead show a scatter in age ranging from $\leq$ 1 Myr to 7 Myr. This type of width in the evolutionary sequence of young clusters is common and is probably due to the combined effect of variable extinction, variability, and/or sources in different evolutionary stages. \\
One interesting feature of Fig. \ref{fig:jjmk} is the potential presence of a group of stars falling on the 5 to 7 Myr isochrones around $J$$\sim$17.5 and with ($J$-\ks) colors in the range 2.0 to 2.4 mag, which could be the signature of an older population. 
To test whether different evolutionary stages are at work in the cluster, we divided the NIR-excess sources into two groups: one located above the 3 Myr isochrone and another located below the 3 Myr isochrone. The spatial distribution of the NIR-excess sources is shown in Fig. \ref{fig:nirex}. 
The main massive stars studied in Roman-Lopez et al. (\cite{rom09}) are also labeled. 
From the spatial distribution of the different YSO populations, it appears that the "red" and probably younger YSOs are mainly distributed around the main cluster region. In Fig. \ref{fig:nirex}, we overplot isophotes from the MSX band D emission at 14.7 $\mu$m (Egan et al. \cite{ega03}), and in the following, we use those contours to spatially divide the sample across the field. Using this spatial separation, the younger sources are also mainly found to be located inside the 14.7 $\mu$m emission contours associated with the cluster, with 85\% of the "red" YSOs located within this contour. The "blue" and potentially older population of YSOs appears to be uniformly distributed over the field, except for the northern region north of the main cluster, where no "blue" YSOs are detected. This distributions seems to indicate the presence of an age gradient when going from Obj2 toward Obj1a and beyond, where the main cluster region (around Obj1a) would be younger than the subcluster (around Obj2) one. This potential age gradient has also been suggested by Santos et al. (\cite{san12}) in their analysis of the cluster evolutionary status. They used the PMS turn-on point as an age indicator and show that most (10 out of 13) of the stars that either have already reached the MS or are currently leaving the PMS stage are positioned closer to (or within) the subcluster. If this is true, then the observed age spread seen in the CMDs could be due to the mixed population of sources at the different ages present in the region.

If we restrict the age estimate to the region located within the boundary of the 14.7 $\mu$m contours (i.e., within 35 arcsec radius or 0.22 pc of the massive Obj1a B0 star), we found that the majority of the sources are less than 3 Myr old. Among these, if we assume that all the sources falling below the 1 Myr isochrone have an age $\leq$ 1 Myr, then the median age of the cluster would be $<$ 1 Myr. It is worth noting that the PMS models are still quite uncertain for ages younger than about 1 Myr (see, e.g., Baraffe et al. \cite{bar02}), and one should be cautious in assigning very young ages to PMS stars. Then, if we consider all the YSOs candidates of the GeMS/GSAOI FoV, our age estimate increases to $\sim$3 Myr, which tends to demonstrate that the main cluster region is indeed younger than the rest of the field. Obj2 is presumably responsible for the ionization of  the evolved HII region (see Sect. \ref{ssec:ioniz}), hence presumably part of the older population of the region. In fact, if consider only the region around Obj2, we estimate an age of 4-5 Myr, but the statistical uncertainties are large, since the number of YSOs is low. There are only 12 YSOs among the 57 detected with {\it Spitzer} that are lying outside of the main-cluster region.\\
Another point that might affect the age estimate is the extinction. So far, we have used a constant value over the field of \av = 8.5 mag. The main cluster region, which is more embedded, might be seen through a higher extinction value, hence artificially biasing the age estimate. To test this possibility, we measured the extinction following the method presented in Sect. \ref{ssec:exc} for the main cluster alone. Doing so, we found a mean value of \av = 9.0 mag, which is very consistent with the extinction of \av = 8.9 derived by Roman-Lopes et al. (\cite{rom09}) from their spectroscopy study. If we apply a higher reddening value, for instance if we use \av = 10 mag instead of 8.5 mag, then the age estimate within the main cluster region increases to $\leq$ 2 Myr, which is still younger than the rest of the field. We therefore conclude that extinction alone cannot explain the age difference seen over the field and that the main part of the cluster, associated with the younger YSOs, is probably very young, with an age $\leq$ 2 Myr.


\section{Luminosity function and cluster IMF}
\label{sec:imf}

Stars in clusters have roughly the same age and metallicity, and they are located at the same distance. In addition, in the case of young clusters, the effects of stellar and dynamical evolution are minimal, then the observed present day mass function is a fair representation of the underlying IMF. In this section we first derive the cluster's $K$-band luminosity function, which we use later to build the IMF.

\subsection{K luminosity function}
\label{ssec:klf}

\begin{figure}
   \centering
     \begin{tabular}{c}
    \includegraphics[width= 0.9\linewidth]{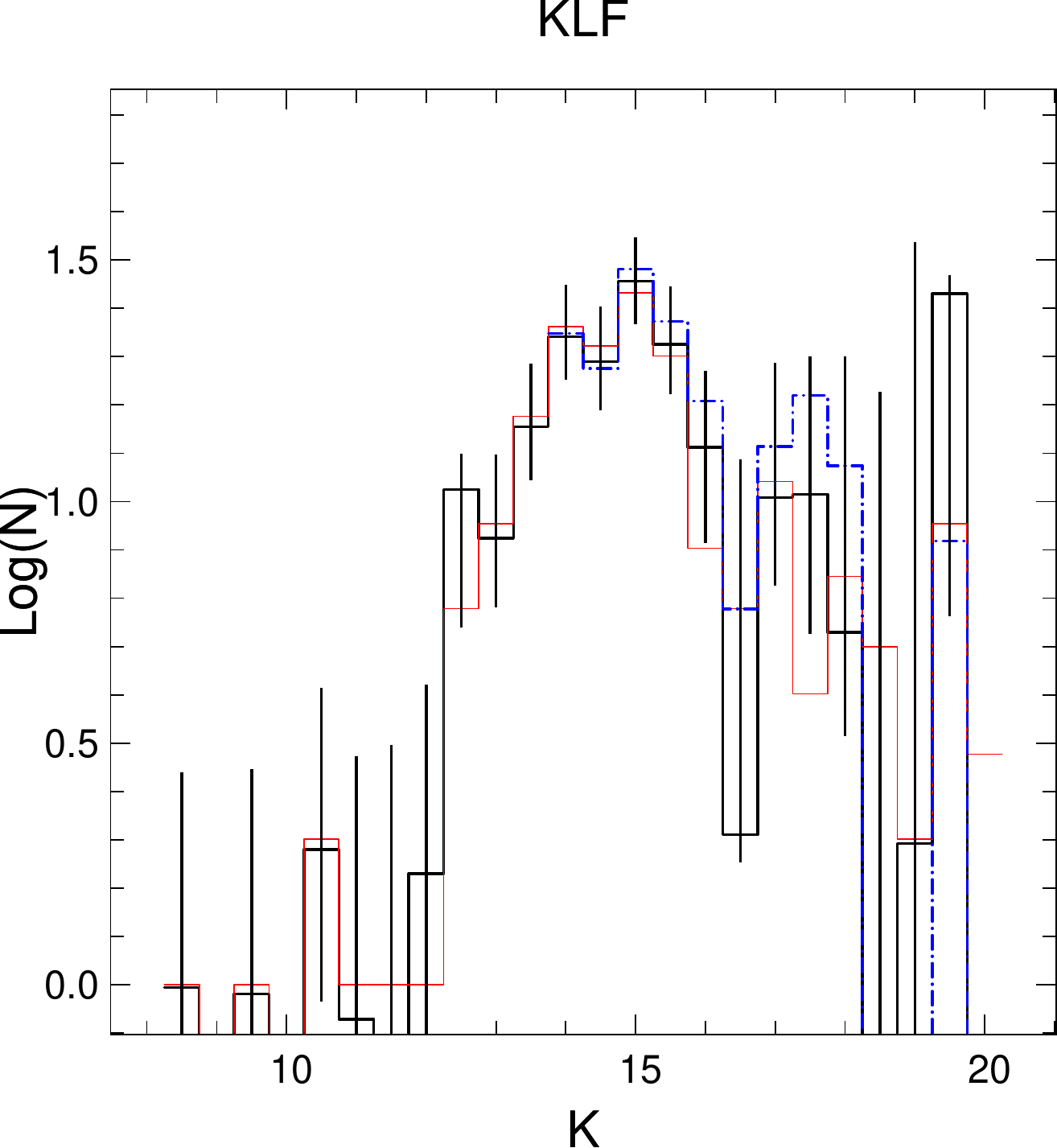}
  \end{tabular}
  \caption{K luminosity function built with different contamination-correction methods. The black histogram shows the KLF when star counts from the galaxy model is used, the blue dash-dot histogram shows the KLF when star count from the control field is used, the red histogram shows the KLF when the color-cut method is used to compensate for contamination (see Sect. \ref{ssec:conta}). Vertical error bars are counting statistics uncertainties computed as $\sqrt N$ for the black histogram.}
              \label{fig:klf}
\end{figure}

To derive the K-band luminosity function (KLF), one first needs to correct for field contamination and incompleteness. Incompleteness is not especially an issue with this data, since the limiting magnitudes are deep. However, contamination is significant, as demonstrated in Sect. \ref{ssec:conta}, and one needs to statistically subtract field stars from cluster members. In Fig. \ref{fig:klf}, we draw the KLF when different field subtraction techniques are used. 
For each case, the correction is done for each bin by subtracting the contamination star count number, normalized to the GeMS/GSAOI field, to the global star count measured on the image. All histograms provide an almost similar KLF, at least within counting statistics uncertainties, which shows that our field contamination correctly removes field stars.

\subsection{IMF}
\label{ssec:imf}
\begin{figure}
   \centering
  \includegraphics[width= 0.9\linewidth]{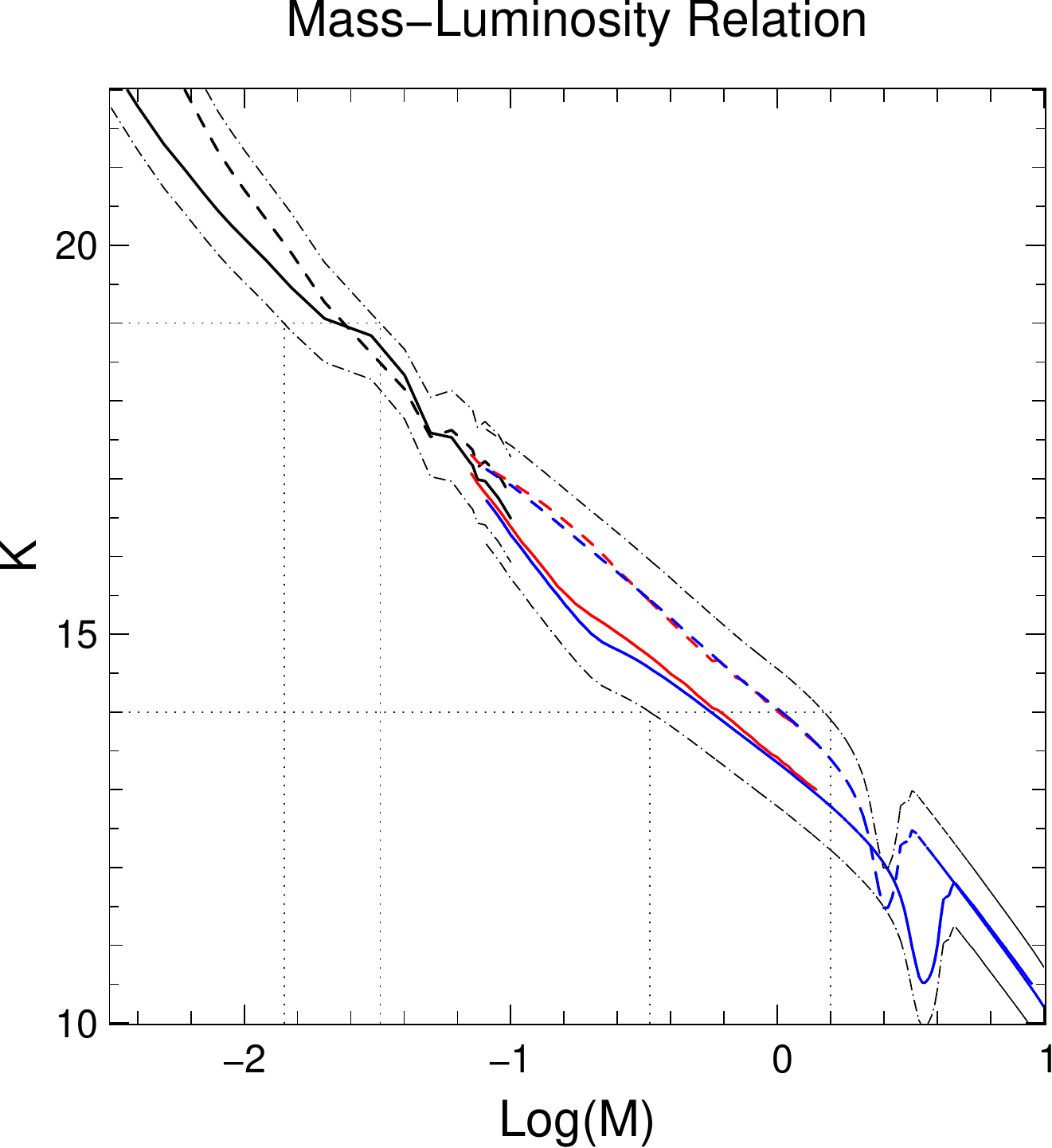} \\
  \caption{Mass luminosity relation. Black lines indicate the Baraffe 03 model, red the Baraffe 98, and blue Tognelli 11. Full lines denote an age of 1 Myr, and dashed lines denote an age of 2.5 Myr. Dash-dotted lines are for when uncertainties on the distance ($\pm$ 0.2 kpc) and on the extinction ($\pm$ 0.2 mag) are summed in quadrature. The dotted lines show, for a given observed magnitude,  the expected errors in mass. Those errors are used in the IMF Fig. \ref{fig:imf2}}
              \label{fig:mlr}
\end{figure}

To convert apparent magnitudes to absolute ones, we consider the distance d = 1.3 $\pm$ 0.2 kpc and a constant foreground extinction of $A_K$ = 1.0 $\pm$ 0.2 mag. The mass of each star should then be estimated based on a mass-luminosity relation (MLR). There are several MLR  available, covering different mass range and ages. Figure \ref{fig:mlr} displays three of them, namely models from Baraffe et al. (\cite{bar03}), Baraffe et al. (\cite{bar98}), and Tognelli et al. (\cite{tog11}). Those two last theoretical models have been converted into magnitudes using the semi-empirical bolometric corrections by Bell et al. (\cite{bel14}). All three models are displayed with two different ages: 1 Myr and 2.5 Myr. 
As described by Muench et al. (\cite{mue00}), different models are generally in good agreement, and the main uncertainties are coming from the age indetermination rather than model discrepancies. Based on the analysis done in Sect. \ref{ssec:age}, we have shown that an age gradient may be present in the field, with a younger population concentrated on the main cluster region. To reduce as much as possible uncertainties coming from this age gradient, in the following we therefore only consider the main cluster region for which an age of 1 $\pm$1 Myr has been estimated. The corresponding 1 Myr MLR will then be considered.

\begin{figure}
   \centering
  \includegraphics[width= 0.9\linewidth]{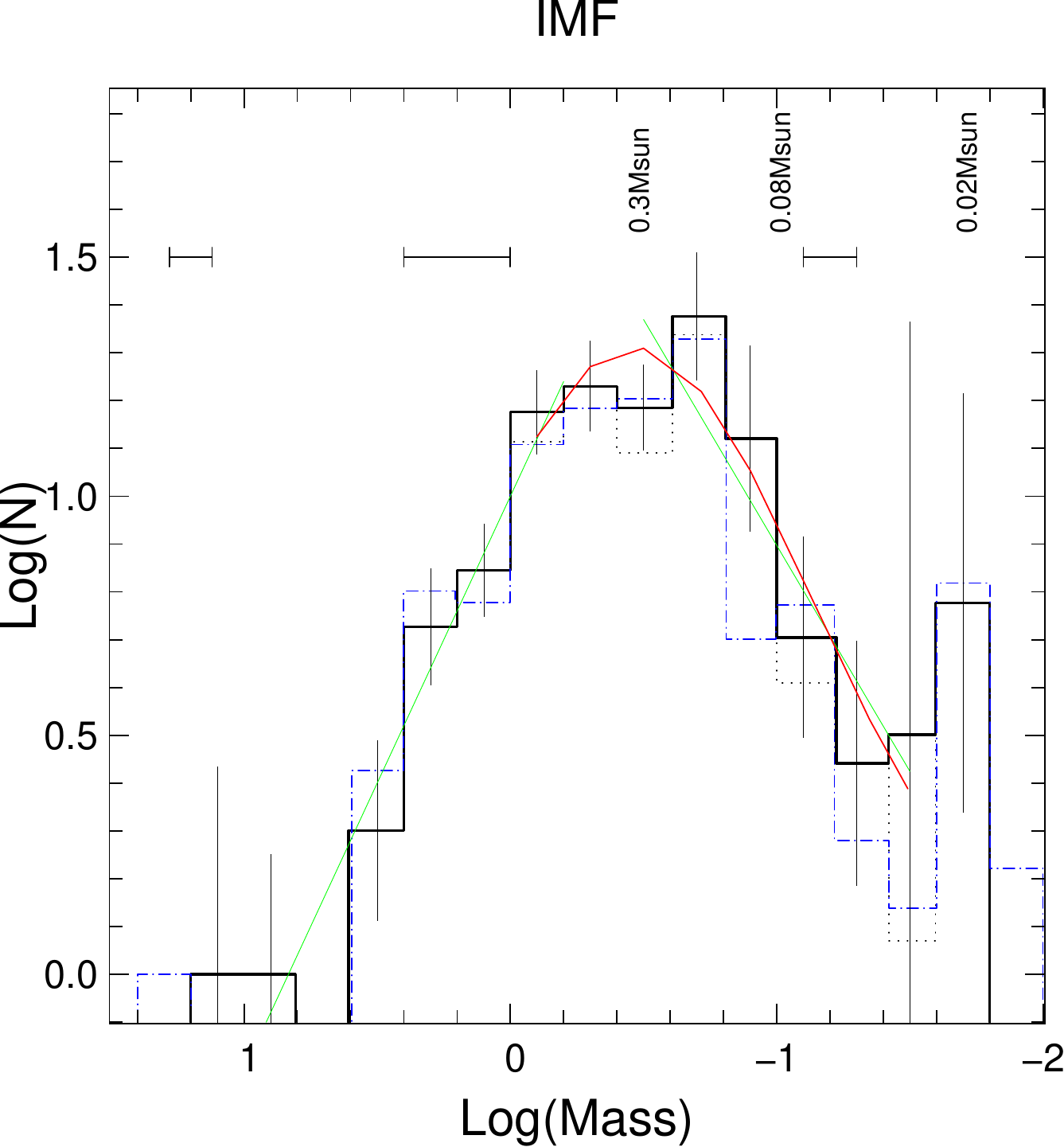} \\
  \caption{IMF derived from the KLF (black solid line). The dotted line shows the IMF derived when NIR excess sources are removed from the sample. The blue dot-dashed line shows the IMF derived from the $H$-band LF. Errors in Log(Mass) are shown as horizontal lines, and they are estimated based on Fig. \ref{fig:mlr}. Vertical lines show the counting statistics error for the black histogram. The slope of the IMF's first part, fitting from the black-full line histogram, gives $\Gamma$ = $-$1.25 $\pm$ 0.2, and is shown in green. The slope of the second part is $\Gamma$ = +0.95 $\pm$ 0.2, if we remove the peak at 0.02 \msun from the fit, and is also shown in green. Chabrier et al. (\cite{cha03}) log-normal like distributions (red full line) gives a critical mass $m_c$=0.35 $\pm$ 0.1 \msun and $\sigma$ = 0.45 $\pm$ 0.1 \msun. The total mass, when integrating over all the bins, is 78 $\pm$ 18 \msun.}
              \label{fig:imf2}
\end{figure}

The IMF for the RCW\,41 main cluster is presented in Fig. \ref{fig:imf2}. 
There are several sources of errors that can affect the IMF shape (e.g., Harayama et al. \cite{har08}), and uncertainties must be treated carefully. 
The principal sources of error in this work are the counting statistics, which are strongly affected by the contamination correction at low magnitudes, hence by low-mass stars and by the age uncertainty, which affects the MLR relation. In addition, estimating the IMF from the $K$-band luminosity function may be biased by the presence of NIR-excess sources. In Sect. \ref{ssec:yso}, we identified 28 NIR-excess sources based on their \jhk colors. 

The number of NIR-excess sources located within the main cluster region is ten. We do not account for the IRAC excess sources, because they only have a negligible excess in the \ks-band luminosity. Once corrected for contamination, the total number of stars is 120, so that NIR excess will convert into a fraction of $\sim$8\%, which is small enough so that it should not affect significantly the IMF. To confirm this result, in Fig. \ref{fig:imf2}, we plot the IMF derived when all NIR-excess sources are removed from the $K$-band sample. Both samples agree very well within the counting and mass uncertainties, and no systematic bias is introduced owing to the inclusion of NIR-excess sources. 
To mitigate the effect of NIR excess sources, Ojha et al. (\cite{ojh09}) used the $J$-band LF instead of the KLF, because it is less affected by circumstellar matter, compared to what is expected solely from the stellar photosphere. We did not try to derive a $J$-band LF, since the $J$-band star-counting numbers might be affected by an additional statistical error (see Sect. \ref{ssec:conta}), but also because we could not apply the color-cut contaminant selection for the $J$-band. Instead, we derived the $H$-band luminosity function, and the corresponding IMF. The $H$-band color should be less affected than the \ks-band by the NIR excess, and the control-field limiting magnitude is also deeper in $H$ than in \ks, providing a better constraint on the low-mass end of the IMF. The IMF derived from the $H$-band LF also agrees very well with the \ks-band one, at least within the counting-statistics error bars, which confirms that the impact of NIR excess sources does not affect the IMF shape significantly in our case. \\
We chose not to include a correction factor for unresolved binary. This choice is motivated by the fact that, with GeMS/GSAOI, we benefit from a high-angular resolution, which means that a significant fraction of the binary systems (those with a physical separation larger than 130 AU) are actually resolved. In their analysis of NGC\,3603, Harayama et al. (\cite{har08}) found that the impact of unresolved binaries on the IMF slope is less than 0.04. There were able to resolve binaries with separations larger than 490 AU; as a result, we can safely state that with the GeMS/GSAOI resolution, the impact of unresolved binaries on the IMF shape is negligible. 
Taking errors on counting statistics and errors on mass estimation both into account, we can derive the slopes of the different IMF regimes. The first regime, covering stars from 10 to 0.1 \msun\, shows the typical rise in number with decreasing mass into subsolar regime, with Salpeter-like power-law slopes. The determination of the slope is sensitive to the mass and counting uncertainties, but also to the bin size (Ma\'iz Apell\'aniz \& \'Ubeda et al. \cite{mai05}) and the mass range chosen. 

We tried different combinations of mass limits and bin selection (respectively 0.05, 0.1 and 0.2 \msun). We also tried non-regular bin sizes following the prescription by Ma\'iz Apell\'aniz \& \'Ubeda et al. (\cite{mai05}). From these experiments, we found a difference in the slope of  $\Delta \Gamma$ = 0.2 dex. Finally, we also compared the slopes measured from the IMF derived from the KLF and HLF, and from the NIR-excess removed sample, but all are giving very similar slopes. Those results are summarized in Table \ref{tabimfparam}. In conclusion, a conservative result would be that the first regime follows a power law with $\Gamma$ = $-$1.25 $\pm$ 0.2. \\
We also fit the subsolar IMF regime, from $\sim$0.4 \msun\, down to 0.03 \msun. In the  subsolar regime, RCW\,41 shows a steep decline toward fainter magnitudes and lower masses. If we fit this region with a power law, we find a slope of $\Gamma$=0.95. If we apply a Chabrier et al. (\cite{cha03}) log-normal like distribution to the subsolar regime, we derive a critical mass $m_c$=0.35 $\pm$ 0.1 \msun\, and $\sigma$ = 0.45 $\pm$ 0.1 \msun. Those values are close, but slightly higher than the canonical ones derived by Chabrier et al. (\cite{cha03}) ($m_c$=0.2 \msun\, and $\sigma$ = 0.55 \msun). When integrating over all the bins, the total mass is 78 $\pm$ 18 \msun. All the results are summarized in Table \ref{tabimfparam}.

\begin{table*}
\caption{Slope and parameters of the different mass functions. }            
\label{tabimfparam} 
\begin{tabular}{ccccc}
\hline
\hline
IMF & High mass slope & Low mass slope & Log-normal fit & Total mass \\
 & 10 $<$ \msun $<$ 0.5 & 0.3 $<$ \msun $<$ 0.03 & 1 $<$ \msun $<$ 0.03 & \\
\hline
Main cluster &  $\Gamma_1$ = $-$1.25 $\pm$ 0.2 & $\Gamma_2$ = +0.95 $\pm$ 0.2 & $m_c$=0.35 $\pm$ 0.1 \msun ; $\sigma$ = 0.45 $\pm$ 0.1 \msun & 78 $\pm$ 18 \msun\\
Main cluster - no NIR excess &  $\Gamma_1$ = $-$1.21 $\pm$ 0.2 & $\Gamma_2$ = +1.1 $\pm$ 0.2 & $m_c$=0.3 $\pm$ 0.1 \msun ; $\sigma$ = 0.4 $\pm$ 0.1 \msun & 75 $\pm$ 21 \msun \\
Main cluster - derived from HLF &  $\Gamma_1$ = $-$1.18 $\pm$ 0.2 & $\Gamma_2$ = +1.2 $\pm$ 0.2 & $m_c$=0.35 $\pm$ 0.1 \msun ; $\sigma$ = 0.4 $\pm$ 0.1 \msun & 77 $\pm$ 20 \msun \\
\hline
\end{tabular}
\end{table*}

\section{Discussion}
\label{sec:discu}

\subsection{Comparison of the RCW\,41 cluster's IMF with other young clusters}
We can compare the IMF derived for the young cluster embedded within RCW\,41 with those of other well studied young clusters, such as IC348 and the Trapezium (Muench et al. \cite{mue03}, Luhman et al. \cite{luh00}), Chamaeleon I (Luhman et al. \cite{luh07}), NGC\,2024 (Levine et al. \cite{lev06}), NGC\,2264 (Sung et al. \cite{sun10}), NGC\,6611 (Oliveira et al. \cite{oli08}), NGC\,7538 (Mallick et al.  \cite{mal14}), W3 Main (Ojha et al. \cite{ojh09}, Bik et al. \cite{bik14}), and NGC\,3603 (Harayama et al. \cite{har08}). \\
As for these clusters, RCW\,41 has an IMF that rise in number with decreasing mass into the subsolar regime, with Salpeter-like power-law slopes. The RCW\,41 slope with $\Gamma$=$-$1.25 is, however, less steep than for IC 348 ($\Gamma$ = $-$1.53) but very consistent with the Trapezium one ($\Gamma$ = $-$1.21) or W3 main ($\Gamma$= $-$1.1). 

Another important feature of the young cluster's IMF is the turning point in mass. From the analysis of Fig. \ref{fig:imf2}, we find that the RCW\,41 IMF peaks at a mass of $\sim$ 0.3 \msun. Such flattening of the IMF around 0.5 \msun\, has also been observed in other clusters (e.g., Bastian et al. \cite{bas10}, Dib S. \cite{dib14}, Luhman et al. \cite{luh07}), although it seems that RCW\,41 is on the higher mass range than other similar young clusters. For instance, the Trapezium, IC 348, Chamaeleon I and NGC\,2024 all have a mode in a range 0.1 to 0.2 \msun. On the other hand, the Taurus cluster (e.g., Luhmann \cite{luh00}, Brice\~no et al. \cite{bri02}) is significantly different, with an IMF that peaks at $\sim$ 0.8 \msun. The properties of RCW\,41 are similar to those of NGC\,6611, for which Oliveira et al. (\cite{oli08}) derived a peak at $\sim$0.5 \msun. Fitting a Chabrier log-normal distribution, they derived a critical mass $m_c$=0.40 $\pm$ 0.04 \msun\, and $\sigma$ = 0.56 $\pm$ 0.04 \msun, which is slightly higher than but consistent with the values derived for RCW\,41 ($m_c$=0.35 $\pm$ 0.1 \msun\, and $\sigma$ = 0.45 $\pm$ 0.1 \msun). It is difficult to explain the differences seen on the characteristic mass for the different clusters. In particular, we should point out that, in the mass range where the IMF peaks, error bars are large because of the differences in the MLR depending on ages. In the case of Taurus, Luhmann (\cite{luh00}) and Brice\~no et al. (\cite{bri02}) interpreted the higher mass as an unusually high average Jeans mass. On the other hand, Elmegreen et al. (\cite{elm08}) show that the thermal Jeans mass only weakly depends on environmental factors such as density, temperature, metallicity, and radiation field. Even if different clusters show significant differences in their IMF (e.g., Dib et al. \cite{dib14}), it is still difficult to conclude on the underlying process causing those differences.

In the substellar regime, RCW\,41 shows a steep decline ($\Gamma$=0.95) toward fainter magnitudes and lower masses. 
Muench et al. (\cite{mue03}) derived a consistent decrease for the Trapezium ($\Gamma$$\sim$0.7), but a much steeper decrease for IC348 ($\Gamma$=2.0). 
One striking element when comparing the IMFs of IC348, the Trapezium, and RCW\,41 is the presence of a secondary peak around 0.02 \msun. 
The presence of a secondary peak in the Trapezium below the hydrogen-burning limit has also been observed by Slesnick et al. (\cite{sle04}), although they find it at a slightly higher mass (0.05 \msun\,  for Slesnick et al. vs. 0.025 \msun\, for Muench et al.). Levine et al. (\cite{lev06}) also report a possible secondary peak at a mass of 0.035 \msun\, for NGC\,2024, even though they notice that the significance of this peak depends on the chosen IMF bin size. For RCW\,41, this region of the IMF is strongly affected by the contamination correction, and counting statistics error are large. Also, because our control-field limiting magnitude is lower than the science field, we used the calibrated galaxy model star counts for this mass range. Even if the control field and galaxy model star counts appear to be in good agreement, we cannot rule out any bias. For instance, Da Rio et al. (\cite{dar12}) analyzed the Orion nebulae IMF, did not find any evidence for a secondary peak, and interpreted this result as the potential impact of an inaccurate background contamination correction in the previous studies. If the significance of this secondary peak is true, one interesting implication of such a structure in the cluster's substellar IMF would be the existence of a separate formation mechanism for very low-mass brown dwarfs. However, as pointed out in Muench et al. (\cite{mue03}) and Lada \& Lada (\cite{lad03}), the significance that should be attached to this feature depends on the accuracy of the adopted MLRs for subsolar objects, and it could also be an artifact introduced by the MLR for brown dwarfs.

\subsection{Brown dwarf fraction}
An alternative way to compare different clusters' low-mass IMF is to evaluate the ratio of brown dwarf to stars. Since our data are complete down to 0.01 \msun, we should get a good estimate of the  brown dwarf fraction for the RCW\,41 cluster. We followed the prescription by Brice\~no et al. (\cite{bri02}), which takes the ratio of the number of stars for a mass range of 0.02 to 0.08 \msun\, divided by the number of stars in a mass range of 0.08 to 10 \msun. 
Doing so, we derived a brown dwarf fraction of 18 $\pm$ 5\%. This number is very consistent with the fraction of 19\% derived by Oliveira et al. (\cite{oli08}) for NGC\,6611 and the 14\% derived for IC348 by Muench et al. (\cite{mue03}). It is also reasonably consistent with the fraction of 22\% derived for the trapezium also by Muench et al. (\cite{mue03}). It is, however, smaller than the 26\% derived by Luhman et al. (\cite{luh07}) for Chamaeleon I, and lower than the fraction of 30\% derived by Levine et al. (\cite{lev06}) for NGC\,2024. Finally, it is significantly lower than the $\sim$43\% derived for NGC\,1333 by Scholz et al. (\cite{scho12}), which is actually the largest fraction observed for young clusters. 
Levine et al. (\cite{lev06}) suggest that the lower fraction seen for IC 348 could be due to the lack of photoionizing stars for that cluster. In fact, isolated brown dwarfs might be formed through the photoevaporation of accretion disks around prestellar cores by late O to early B main-sequence star as suggested by models (e.g., Whitworth \cite{whi04}, Robberto et al. \cite{rob04}). The Trapezium, NGC\,2024, and Chamaeleon I do have the presence of O and/or early B stars and for those clusters, the photoevaporation scenario could explain the higher brown dwarfs' fraction.
In the case of RCW\,41, the presence of late O (Obj2) and early B (Obj1a) stars has been spectroscopically confirmed. This is also true for NGC\,6611 (Oliveira et al. \cite{oli08}), where the cluster hosts O3 and O5 stars. However, both clusters have a smaller brown dwarf fraction. For NGC\,6611, Oliveira et al. (\cite{oli08}) suggest an alternative process based on gravitational fragmentation of infalling gas, giving rise to filamentary-like structures within which lower-mass clumps would form (e.g. Bonnell et al. \cite{bon08}, Thies et al. \cite{thi15}). The presence of filamentary-like structures (see Fig. \ref{FigH2}), converging toward the NIR cluster of RCW\,41, could be an indication that such a process is also at work in the formation of the brown dwarfs in RCW\,41. As pointed in Luhman et al. (\cite{luh07}), it is important to highlight that small systematic offsets in mass estimates can result in large differences in the relative numbers of stars and brown dwarfs, so that looking for accurate differences between the different clusters might be difficult. In any case, we can conclude that the brown dwarf population represents about one in the four of the RCW\,41 cluster members, which is consistent with other nearby young clusters (e.g. see Kroupa et al. \cite{kro13} for a review). 


\subsection{RCW\,41 ionizing source}
\label{ssec:ioniz}
The exciting star of the \HII region has not been clearly identified so far. According to Ortiz et al. (\cite{ort07}), Obj1a (B0 v) and Obj2 (O9 v) are the best candidates for ionization sources of the RCW~41 region. At 5.8 $\mu$m, the  region shows a cavity at its center, surrounded by PAH emission in the PhotoDissociation Region (PDR). The  H$\alpha$ emission is strongest in the center of the nebula (see Fig. \ref{FigH2}), implying that the ionizing source might be embedded within the nebula. This phenomenon has been observed in many bubbles associated with \HII regions (e.g., Deharveng et al. \cite{deh10}). Thus, we believe that the exciting star must be inside the 5.8 $\mu$m cavity that surrounds the H$\alpha$ emission, which excludes the Obj1a source as a potential candidate. Also Obj1a is at the extreme edge of the  H$\alpha$ emission, which makes it unlikely to be the ionizing candidate. To identify the probable ionizing candidates of the individual \HII regions, we then used a \jhk catalog of the region inside the cavity and searched for ionizing sources in a circular area of radius $\sim$ 1.15 pc (i.e., the radius of the 5.8 $\mu$m cavity) from the center of the bubble. Using the extinction laws of Rieke \& Lebofsky (\cite{rie85}), the observed ($J$-$H$) color and M$_J$-spectral type calibration table of Bessel \& Brett (\cite{bes88}), and a distance of 1.3 kpc, we selected luminous sources that have spectral types earlier than B3 MS star. We followed the same prescription as described in Samal et al. (\cite{sam10}) of rejecting the most likely giants and foreground sources based on ($J$ $\times$ [$J-H$]) and ([$J-H$] $\times$ [$H-Ks$]) diagrams. After these eliminations, we are left with only one massive O-type star located close to the northwestern border of 5.8 $\mu$m cavity. This source is associated with a group of stars in its close vicinity, possibly part of an exciting cluster. The source corresponds to the "Obj2" identified by Santos et al. (\cite{san12}), whose spectroscopic spectral type agrees with the photometric spectral derived form our observations. From the above discussion, we conclude that the Obj2 source is the most likely ionizing star of the RCW\,41 \HII region.

\subsection{Star formation toward RCW\,41}
\begin{figure}[ht!]
   \centering
  \includegraphics[width= 0.99\linewidth]{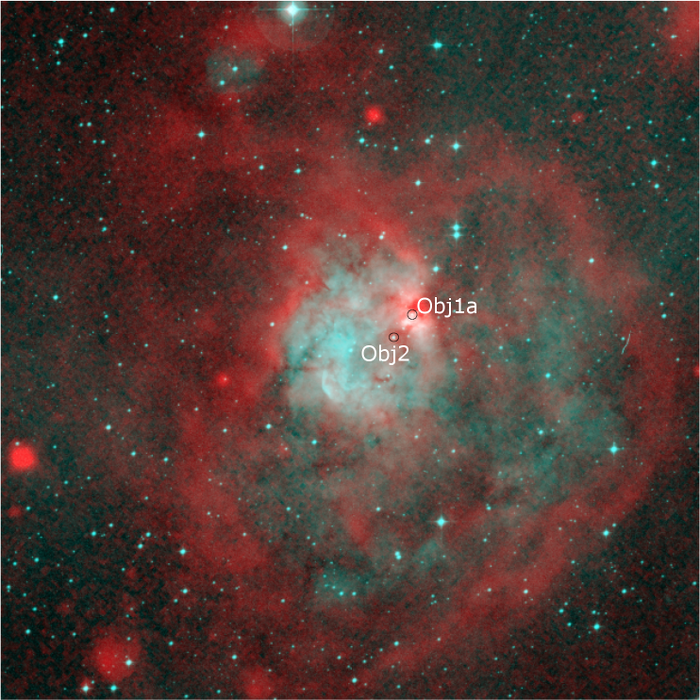}
   \caption{Color composite image of RCW41: H$\alpha$ emission from the UKST is in turquoise, MSX 8.2 micron (band A) emission in red. The field is 28\arcmin $\times$ 28\arcmin. North is up, east is left. The position of Obj1a and Obj2 is also shown.}
              \label{fig:msx}
\end{figure}

Santos et al. (2012) studied the optical polarization toward the young cluster. Their Fig. 11 shows a color-composite image of RCW41 showing the optical emission from the ionized region in blue and the Midcourse Space eXperiment (MSX) band A emission. MSX band A is centered at 8.2 micron and shows the emission of polycyclic aromatic hydrocarbons (PAHs) molecules. Distribution of the $R$-band polarization vector is superimposed on their Fig. 11. For clarity we present in Fig. \ref{fig:msx} the same color composite image: the H$\alpha$ emission from the SuperCOSMOS H$\alpha$ survey (Parker et al. \cite{park05}) is shown, along with the MSX band A. The optical ionized region is composed of two parts: a central part with bright emission and a more diffuse emission located in the western part. The MSX band-A emission delineates the external bright part of the ionized region (see Fig. \ref{fig:msx}) outlining the material (gas and dust) that has been collected during the expansion of the ionized gas between the ionized and the shock fronts (see Elmegreen \& Lada \cite{elm77}). Part of this bright MSX 8.2 micron emission is organized in filaments that converge at the center of the young cluster. The convergence of those filaments has to be proven with dynamical molecular data that are not available yet. However, converging flows are an important physical process in the formation of high-mass stars (see Kumar et al. \cite{kum07}, Schneider et al. \cite{schn10}, \cite{schn12}) and should be considered as a possible mechanism for the formation of the clump at the origin of this bright cluster formation and/or to the possible active continuous accretion of molecular material that can feed the cluster and allow formation of high-mass stars. 
Molecular-line observations are needed to further address what triggers the formation of the clump and how the filaments are active in the accretion process. Another question that has to be addressed is whether the star formation of the bright cluster has been triggered by the interaction with the expansion of the ionized region or not. A gradient of age is observed across the cluster, the bluest and likely ionizing sources of the region being observed on the lower part of the cluster, whereas young embedded sources are observed in the upper part, and a dense HCNO clump is observed farther away (see Fig. 1b in Ortiz et al. \cite{ort07}). Active high-mass star formation is ongoing in this dense clump (see Hill et al. \cite{hil05}, \cite{hil09}). This suggests that star formation progresses toward the clump and could have been triggered by the interaction of the ionized region with the clump. Complementary data are needed to confirm or disprove this scenario.

A more diffuse optical emission extends toward the west of the bright central part and is delineated by fainter MSX 8.2 micron emission, both filamentary and circular, that surrounds the external part of the ionized region. We suggest that the ionizing flux from the exiting stars leaks from the central part and propagates into the surrounding medium as observed in other bubble \HII regions (Deharveng et al. \cite{deh09}). The outer part of the ionized region corresponds well to the photodissociation region observed at 8.2 micron. Leaking of the ionized emission toward the southwest part of the region suggests that the ionized region is open in this direction (lower density region) and that the ionized gas is coming toward the observed. We suggest that the ionized region is opened toward the observer and is viewed with an angle (beween 0 - face on and 90- edge on) on the line of sight. The young forming cluster appears as a bright spot of 8.2 micron emission observed {\it{\emph{in the direction of}}} the ionized gas. We suggest that the cluster is forming {\it{\emph{at the surface}}} of the ionized region due to the collapse of a clump that interacts with the ionized gas from the expanding \HII region. The origin of the clump formation (pre-existing, converging flow, collapse from the dense surrounding layer of collected material) cannot be inferred from the existing data. 


\section{Summary and conclusions}
\label{sec:conclu}
In this paper we have presented deep, high-angular resolution images of a NIR cluster located at the edge of the RCW\,41 \HII region. Those images have been used to explore the stellar content of the cluster, as well as the star formation history of the region. Based on these observations, our results can be summarized as follows:
 \begin{enumerate}
  \item We revealed an impressive performance of the GeMS/GSAOI MCAO system over a large FoV.  As part of the technical analysis, we derived that the averaged resolution obtained with the GeMS/GSAOI images were 150 mas in $J$, 140 mas in $H$ and 95 mas in \ks, with variations of less than 20\% over a field of almost 100 \arcsec accros. This is a factor of 5 to 8 times better than what can be achieved with seeing-limited instruments, and it shows a significant improvement in the spatial stability of the MCAO performance when compared to SCAO systems. This unique performance demonstrates the opportunities offered by MCAO systems for dense regions like young clusters, in particular to study objects at distances unachievable before. We believe that efforts should be devoted to developing new analysis tools, specific for MCAO systems, that would reduce the photometry uncertainties and optimize the scientific return. 
   \item Using IRAC observations to complement our deep \jhk images, we were able to identify a total of 80 YSOs. Using CMDs and PMS isochrones, we derived the age of these YSOs in the range $<$1 Myr to 5 Myr. More precisely, one-third of the YSOs exhibit an age between 3 Myr and 5 Myr, while two-thirds are $\leq$3 Myr. Looking at the spatial distribution of these two populations, we show a potential gradient over the field, with the oldest YSOs mostly concentrated toward the subcluster region (southeast), while the youngest YSOs are mainly distributed around the main cluster center (northwest). This age gradient suggests sequential star formation, with the youngest part potentially being triggered by the first massive star generation. 
    \item The exquisite resolution delivered by GeMS/GSAOI allowed us to characterize the stellar mass distribution down to 0.01 \msun (which is $\sim$10 M$_{Jup.}$) and to build the IMF of the stellar cluster. Based on the IMF, we derived a total cluster mass of $\sim$78 $\pm$ 18 \msun. The comparison of the RCW\,41 cluster IMF with other nearby young clusters shows that it resembles the Trapezium or IC\,348 clusters in terms of different regimes, with a Salpeter-like power law at high-to-intermediate mass, a flattening around 0.3 \msun, and a steep decrease into the subsolar mass range. The high-to-intermediate mass regime is well fit by a $\Gamma$= -1.25 $\pm$ 0.2 slope. The IMF peak appears to be shifted toward high masses when compared to Trapezium or IC\,348, and it is more consistent with NGC\,661. For the very low-mass regime, we derived a brown-dwarf fraction of 18 $\pm$ 5\%, which is consistent with other nearby young clusters like Trapezium or Chamaleon I, but significantly higher than IC\, 348. Taken together, these results suggest that the medium-to-low mass end of the IMF depends on environment.
   \end{enumerate}



\begin{acknowledgements}
Based on observations obtained at the Gemini Observatory, which is operated by the Association of Universities for Research in Astronomy, Inc., under a cooperative agreement with the NSF on behalf of the Gemini partnership: the National Science Foundation (United States), the National Research Council
(Canada), CONICYT (Chile), the Australian Research Council (Australia), Minist\'erio da Ci$\rm{\hat{e}}$ncia, Tecnologia
e Inova{\c{c}}$\rm{\tilde{a}}$o (Brazil) and Ministerio de Ciencia, Tecnologiaa e Innovaci\'on Productiva (Argentine).\\ 
We thank the GeMS/GSAOI commissioning team, namely Rodrigo Carrasco, Peter Pessev, Fabrice Vidal, Claudia Winge, for their efforts to collect the data. We also thank the anonymous referee for the comments thath certainly improved the quality of the paper.\\
B. Neichel and A. Bernard acknowledge the financial support from the French ANR program WASABI to carry out this work.\\
H. Plana thanks the LAM scientific council for its financial support during the staying at LAM in december 2013 and the CNPq/CAPES for its financial support using the PROCAD project 552236/2011-0.\\
M. R. Samal acknowledges the financial support provided by the French Space Agency (CNES) for his post-doctoral fellowship. 
\end{acknowledgements}


\end{document}